\documentclass[aps,prb,reprint,twocolumn,superscriptaddress]{revtex4-2}

\usepackage{graphicx}
\usepackage{bm}
\usepackage{hyperref}
\usepackage{amsmath} 
\usepackage{amssymb}
\usepackage[table]{xcolor}
\usepackage{array}
\usepackage{multirow}
\usepackage{physics}
\usepackage{lipsum}

\begin{document}
\title{Fractional quantum anomalous Hall states in twisted 
bilayer MoTe$_2$ and WSe$_2$}
\author{Aidan P. Reddy}
\email{areddy@mit.edu}
\affiliation{Department of Physics, Massachusetts Institute of Technology, Cambridge, Massachusetts 02139, USA}
\author{Faisal Alsallom}
\affiliation{Department of Physics, Massachusetts Institute of Technology, Cambridge, Massachusetts 02139, USA}
\author{Yang Zhang}
\affiliation{Department of Physics and Astronomy, University of Tennessee, Knoxville, Tennessee 37996, USA}
\affiliation{Min H. Kao Department of Electrical Engineering and Computer Science, University of Tennessee, Knoxville, Tennessee 37996, USA}
\author{Trithep Devakul}
\affiliation{Department of Physics, Massachusetts Institute of Technology, Cambridge, Massachusetts 02139, USA}
\author{Liang Fu}
\email{liangfu@mit.edu}
\affiliation{Department of Physics, Massachusetts Institute of Technology, Cambridge, Massachusetts 02139, USA}
\date{\today}
\begin{abstract}
We demonstrate via exact diagonalization that AA-stacked TMD homobilayers host fractional quantum anomalous Hall (FQAH) states with 
fractionally quantized Hall conductance at fractional fillings $n=\frac{1}{3},\, \frac{2}{3}$ and zero magnetic field. 
While both states are most robust at angles near $\theta\approx 2^{\circ}$, the $n=\frac{1}{3}$ state gives way to a charge density wave with increasing twist angle whereas the $n=\frac{2}{3}$ state survives across a much broader range of twist angles. We show that the competition between FQAH states and charge density wave or metallic phases is primarily controlled by the wavefunctions and dispersion of the underlying Chern band, respectively. Additionally, Ising ferromagnetism is found across a broad range of fillings where the system is insulating or metallic alike. The spin gap is enhanced at filling fractions where integer and fractional quantum anomalous Hall states are formed. 
\end{abstract}

\maketitle

 The discovery of the integer and fractional quantum Hall effects (QHE) in two-dimensional electron systems under a magnetic field ushered in the paradigm of topological matter and electron fractionalization \cite{klitzing1980new, tsui1982two} over forty years ago. It was recognized shortly thereafter that, while broken time reversal symmetry is a necessary condition for QHE, Landau levels are not: a Bloch band with a nonzero Chern number suffices \cite{thouless1982quantized, haldane1988model}. The possibility of quantum Hall analogs in which time reversal symmetry is broken \emph{spontaneously} at zero magnetic field is a subject of fundamental importance and long-standing interest. Advances in quantum materials have brought the search for such phases, known collectively as quantum anomalous Hall (QAH) states, to the forefront of condensed matter physics.

Following theoretical proposals, transport measurements have demonstrated the integer QAH effect in a variety of material systems \cite{chang2023colloquium}, including magnetically doped topological insulator \cite{chang2013experimental}, intrinsic magnetic topological insulator MnBi$_2$Te$_4$ \cite{deng2020quantum}, magic-angle twisted bilayer graphene \cite{serlin2020intrinsic}, and transition metal dichalcogenide heterobilayer MoTe$_2$/WSe$_2$ \cite{li2021quantum}. Beyond a proof of principle, the experimental demonstration of QHE at zero magnetic field opens a new path to microwave circulators \cite{mahoney2017zero}, chiral topological superconductivity, and Majorana fermion \cite{fu2009probing, akhmerov2009electrically}.  

Even more exciting is the fractional quantum anomalous Hall (FQAH) state, a new 
phase of matter that exhibits fractionally quantized Hall conductance and hosts fractional quasiparticles (anyons) at zero magnetic field. 
Physical realization of FQAH states relies on the synergy between band topology, strong correlation, and spontaneous time reversal symmetry breaking. These states can host new types of fractionalization unseen before in Landau levels. Moreover, proximity coupling between FQAH states and superconductors at zero magnetic field may provide a promising route to topological quantum computing \cite{lindner2012fractionalizing, mong2014universal}. 

Magic-angle twisted bilayer graphene is a theoretically interesting candidate platform for the FQAH state \cite{repellin2020chern, ledwith2020fractional, wilhelm2021interplay, parker2021field, ledwith2022vortexability}. Local compressibility measurements demonstrate its fractional quantum Hall states at magnetic fields above $5$ T \cite{xie2021fractional}. However, at zero and small fields, the incompressible states at fractional fillings are observed to be topologically trivial. 

Recently, a new moiré system, AA-stacked twisted homobilayer of transition metal dichalcogenide (TMD) semiconductors WSe$_2$ or MoTe$_2$, has also been predicted to host FQAH \cite{devakul2021magic, li2021spontaneous, crepel2022anomalous}. Here, narrow moir\'e bands are formed at small twist angles and acquire nontrivial topology from the layer pseudospin structure of their Bloch wavefunctions \cite{wu2019topological}. In addition to band topology and narrow bandwidth, strong atomic spin-orbit coupling in moiré TMDs results in the locking of electrons' spin to their valley degree of freedom. This makes spontaneous Ising ferromagnetism possible at finite temperature, 
a key requisite for the realization of FQAH states \cite{crepel2022anomalous}. 

Very recently, signatures of integer and fractional QAH states in optical measurements of twisted MoTe$_2$ bilayers have been reported \cite{cai2023signatures}. 
Photoluminescence spectra 
clearly show a reduction in intensity and a blue shift in peak energy at integer and fractional fillings of the moir\'e unit cell $n=-1$ and $-\frac{2}{3}$, indicating the emergence of correlated insulators. Furthermore, magnetic circular dichroism measurements reveal robust ferromagnetism over a wide range of hole fillings $0.4 \lessapprox |n| \lessapprox 1.2$ 
The coercive field determined from magnetic hysteresis 
is distinctively enhanced at $n=-1$ and $-\frac{2}{3}$. Remarkably, a linear shift in the carrier densities of the optically detected $n=-1$ and $-\frac{2}{3}$ states with the applied magnetic field is found, with a slope $\frac{\partial n}{\partial B}$ in units of $\frac{e}{hc}$ 
matching $C=-1$ and $-\frac{2}{3}$ respectively, as expected from Streda formula $\frac{\partial{n}}{\partial_{B}}|_\mu = \frac{\sigma_H}{ec}$ for states with integer and fractionally quantized Hall conductance $\sigma_{H} = C \frac{e^2}{h}$. Importantly, the linear dispersion persists down to zero magnetic field. These observations taken altogether provide clear, strong evidence for integer and fractional QAH in hole-doped twisted bilayer MoTe$_2$. 

In an independent experiment around the same time, integer QAH states were observed in twisted bilayer WSe$_2$ at hole fillings $n=-1$ and $n=-3$ by electronic compressibility measurements \cite{foutty2023mapping}. Here, the linear shift in the density of the incompressible state reveals states with quantized Hall conductance $C=+1$, which persist down to zero magnetic field. The topological gap of the QAH state at $n=-1$ is found to be around $1$ meV. 

The discovery of integer and fractional QAH states in twisted bilayer MoTe$_2$ and WSe$_2$ following theoretical prediction \cite{devakul2021magic, li2021spontaneous, crepel2022anomalous} demonstrates the extraordinary richness at the intersection of band topology and electron correlation. 
Many open questions remain to be answered. While prior theoretical studies of FQAH in twisted TMD homobilayers have focused on the filling factor $n=-\frac{1}{3}$ 
and at ultrasmall twist angles $\theta \lessapprox 1.5^{\circ}$ \cite{li2021spontaneous, crepel2022anomalous, morales2023pressure}, the newly observed FQAH state in twisted MoTe$_2$ occurs at the filling fraction $n=-\frac{2}{3}$ in a device with a larger twist angle $\theta=3.7^{\circ}$. In addition, a weak feature indicative of another FQAH state at $n=-\frac{3}{5}$ was observed. 


In this work, we study ferromagnetism, FQAH, and competing states in AA stacked TMD homobilayers.
We begin with a detailed discussion of the system's single particle physics, which evolves dramatically with twist angle. 
We present original \textit{ab initio} calculations for $t$MoTe$_2$ band structure. Next, we demonstrate robust Ising ferromagnetism across a broad range of carrier densities in the lowest moiré band, 
independent of whether the system is metallic or insulating at a given carrier density. For a range of twist angles and realistic Coulomb interaction, we demonstrate FQAH states at filling factors $n=\frac{1}{3}, \frac{2}{3}$. These states are fully spin/valley polarized and exhibit fractionally quantized Hall conductance at zero magnetic field. We find for both fillings that the FQAH gap is largest near $\theta\approx 2^{\circ}$. As twist angle increases, the $n=\frac{1}{3}$ state gives way to a charge density wave (CDW) while the $n=\frac{2}{3}$ state survives to significantly higher twist angles, ultimately giving way to a metal. We find that the angle at which the FQAH-CDW transition for $n=\frac{1}{3}$ occurs is only weakly dependent on interaction strength, whereas that at which the FQAH-metal transition occurs for $n=\frac{2}{3}$ is strongly interaction-strength-dependent. This suggests that the former is controlled primarily by a change in the wavefunctions of the lowest-band Bloch states whereas the latter is controlled primarily by a change in their dispersion. 
within the single-band-projected model, we find that the suppression of charge density wave at $n=\frac{2}{3}$ compared to $n=\frac{1}{3}$ is due to interaction-induced renormalization of band dispersion.


\begin{figure}
  \centering
\includegraphics[width=0.8\columnwidth]{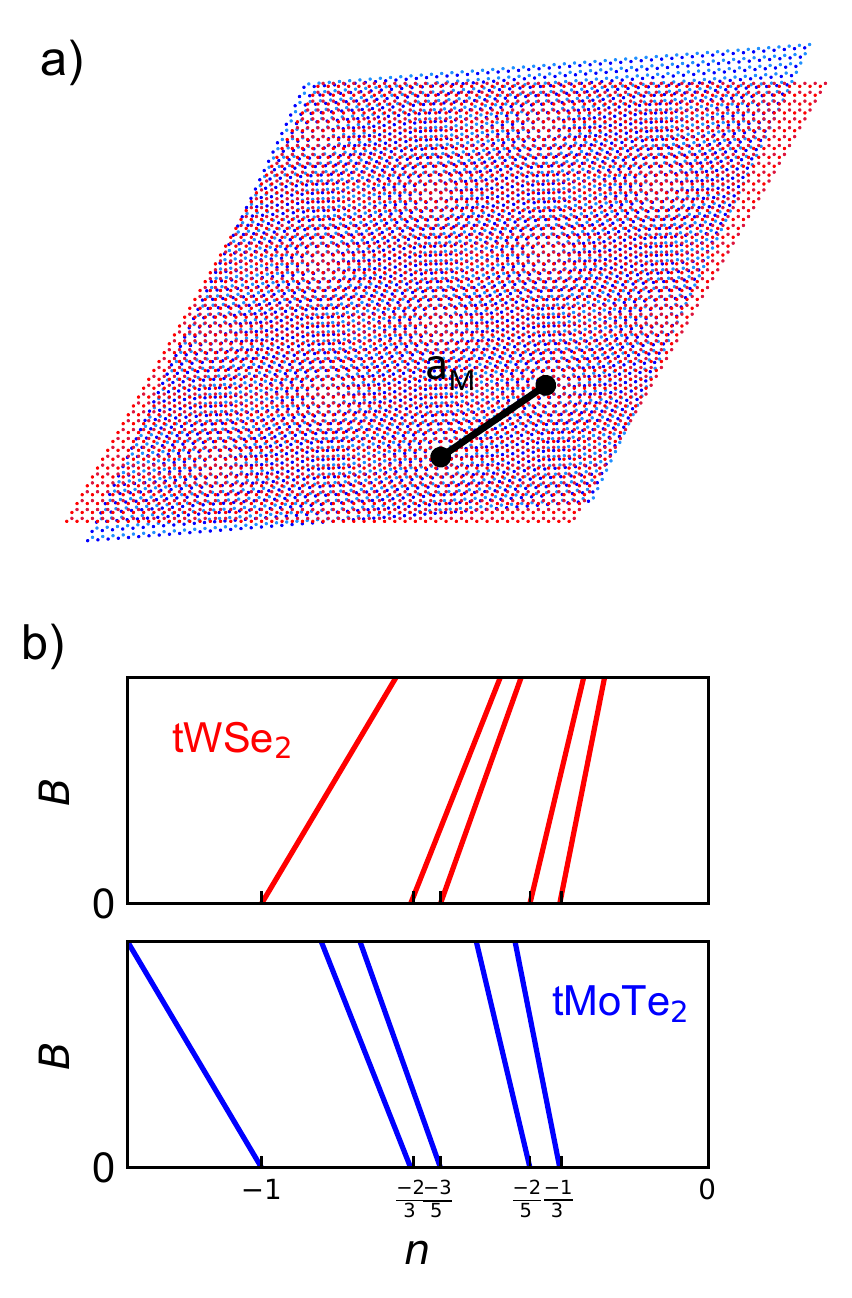}
  \caption{(a) Schematic of AA-stacked homobilayer moiré superlattice and (b) Wannier diagrams showing quantum anomalous Hall states in $t$MoTe$_2$ and $t$WSe$_2$, which have opposite Chern numbers in a given valley.} 
\label{fig:intro}
\end{figure}

\section{Topological moir\'e bands}
\emph{Continuum model --} The valence band edges of a TMD monolayer are located at the $K$ and $K'$ points of the Brillouin zone and have a large effective mass in the range of $0.5-1$ $m_e$. Due to strong atomic spin-orbit coupling, holes at $K$ and $K'$ have opposite spin so that spin and valley degrees of freedom lock into a single two-component ``spin" degree of freedom. When two identical $K$-valley TMD layers 
are stacked with $0^{\circ}$ alignment (AA stacking), holes at a given valley have the same spin in both layers and therefore direct spin-conserving, intra-valley, inter-layer tunneling is present.



Rotation between the two layers modifies the dispersion of low-energy holes by introducing an intra-layer superlattice potential 
and inter-layer tunneling 
that vary spatially with the superlattice periodicity. 
As shown by Wu {\it et al} \cite{wu2019topological}, the continuum model Hamiltonian 
for the spin-$\uparrow$ component takes the form of a $2\times2$ matrix in layer space:

\begin{align}\label{eq:CMHam}
  \mathcal{H}_{\uparrow} = \begin{pmatrix}
  \frac{\hbar^2(- i \nabla - \kappa_+)^2}{2m} + V_1(\bm{r}) & t(\bm{r}) \\
  t^{\dag}(\bm{r}) & \frac{\hbar^2( - i\nabla - \kappa_-)^2}{2m} + V_2(\bm{r})
  \end{pmatrix}
\end{align}

with $\mathcal{H}_{\downarrow}$ its time reversal conjugate. Note that this single-particle Hamiltonian is defined with charge neutrality as its vacuum. 
Accordingly, we have chosen a natural convention for $\mathcal{H}$ to write the continuum model Hamiltonian in terms of the hole operators directly: 
\begin{eqnarray}
H_0 = \sum_{\sigma=\uparrow, \downarrow} \int d {\bm r} \; \psi^\dagger_{\sigma} \mathcal{H}_{\sigma} \psi_{\sigma}, \end{eqnarray} 
where $\psi^\dag_\sigma$ creates a hole in the valence band. As such, the single-particle energy spectrum of $H_0$ is bounded from below. 
 We also define $n$ to be the number of holes per unit cell relative to charge neutrality so that $n$ is positive, opposite to the definition commonly used in experiments. 

 Here, the kinetic energy of holes in a given layer follows a quadratic energy-momentum dispersion centered about its $K$ point. The $K$ points of the two layers are displaced 
 due to the interlayer twist and fold into the corners of the moiré Brillouin zone, $\kappa_{+}$ and $\kappa_{-}$.
 We choose our moiré reciprocal lattice vectors to be $\bm{g}_i=\frac{4\pi}{\sqrt{3}a_M}(\cos\frac{\pi(i-1)}{3},\sin\frac{\pi(i-1)}{3})$ and $\kappa_- = \frac{\bm g_1 +\bm g_6}{3}$, $\kappa_+ = \frac{\bm g_1 +\bm g_2}{3}$. Here $a_M=\frac{a_0}{2\sin(\theta/2)}\approx \frac{a_0}{\theta}$ where $a_0$ is the atomic lattice constant.

The parameters of the moiré potential $V_l(\bm{r}),\, t(\bm{r})$ 
can be fitted to first-principles density functional theory calculations given symmetry constraints that we now discuss.
The most general Fourier expansion of the intra-layer potential to the lowest harmonic is $V_l(\bm{r})=-\sum_{i=1}^6 V_{\bm g_i l}e^{i\phi_{\bm g_i l}}e^{i\bm{g}_i\cdot\bm r}$ where $V_{\bm{g}_i}$ is real and the reality of $V_l(\bm{r})$ requires $\phi_{\bm g_i l} = -\phi_{-\bm g_i l}$. 
It follows from $C_{3z}$ symmetry that
\begin{align}
    V_l(\bm{r})=-2V\sum_{i=1,3,5}\cos(\bm{g}_i\cdot\bm{r}+\phi_l).
\end{align}
Here, the origin of $\bm r$ is defined to be at the center of an MM stacking region. Additionally, symmetry under a twofold rotation that interchanges the two layers 
requires $V_l(\bm{r})=V_{\bar{l}}(-\bm{r})$ and, in turn, 
$\phi_2=-\phi_1 \equiv \phi$. 
The same symmetry consideration also applies to the inter-layer tunneling term, which must take the general form
\begin{align}
  t(\bm{r}) = w(1+e^{i\bm{g}_2\cdot\bm{r}}+e^{i\bm{g}_3\cdot{\bm{r}}}).
\end{align}
This model Hamiltonian has spin $U(1)$ symmetry $([S_z, \mathcal{H}]=0)$, but \emph{not} spin $SU(2)$ symmetry, a property that we will see enables robust Ising type ferromagnetism. 

\maketitle
\begin{figure}[t]
\includegraphics[width=\columnwidth]{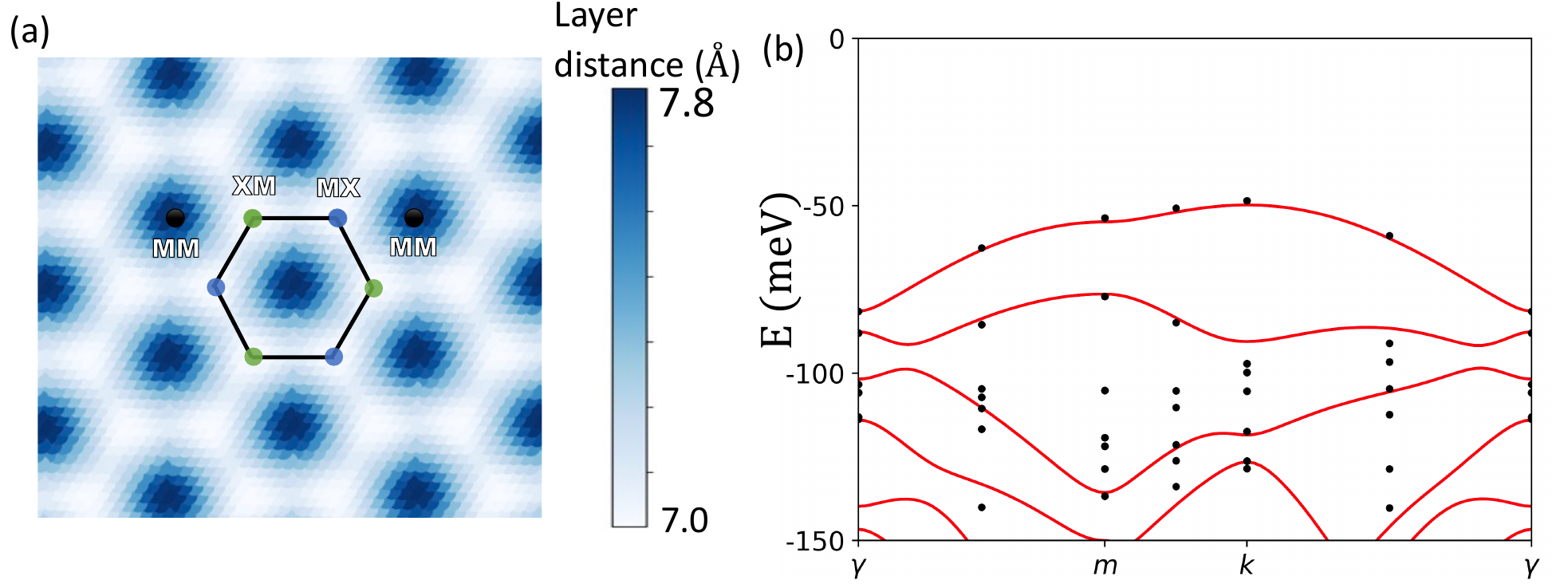}
\caption{
$a)$ The interlayer distance of the twisted MoTe$_2$ structure obtained from DFT is shown, demonstrating a large variation between the MM and XM/MX regions.
$b)$ The continuum band structure (blue lines) is plotted in comparison with large scale DFT calculations (black dots) at twist angle $\theta=4.4^\circ$, showing excellent agreement. Note that additional bands in DFT calculations come from  $\Gamma$ valley states. 
}\label{fig:bzdft}
\end{figure}

\emph{First principles calculations --} 
We now compare the moir\'e band structure of the continuum model with first-principles calculations on twisted bilayer MoTe$_2$. We perform large-scale density functional theory (DFT) calculations with the SCAN density functional ~\cite{peng2016versatile} with dDsC dispersion correction method, which captures the intermediate-range vdW interaction through its semilocal exchange term. 
We find that lattice relaxation has a dramatic effect on moir\'e bands. 
Our DFT calculations at $\theta=4.4^\circ$ with 1014 atoms per unit cell show that the layer distance $d$ varies significantly in different regions of the moir\'e superlattice, as shown in Fig~\ref{fig:bzdft}(a). $d=7.0$\AA{} is smallest in MX and XM stacking regions, where the metal atom on top layer is aligned with chalcogen atom on the bottom layer and vice versa, while $d=7.8$\AA{} is largest in MM region where metal atoms of both layers are aligned. With the fully relaxed structure, the low-energy moir\'e valence bands of twisted bilayer MoTe$_2$ are found to come from the $\pm K$ valley (shown in Fig.1b).


In Fig~\ref{fig:bzdft}c, we compare the band structures of twisted bilayer MoTe$_2$ at $\theta=4.4^\circ$ computed by large-scale DFT and by the continuum model. Remarkably, the low-energy part of DFT band structure is well fitted with the continuum model band structure with parameters stated in Table \ref{paramtable}. 
Importantly, our direct large-scale DFT calculation reveals a significantly narrower moir\'e bandwidth than reported in the previous model study \cite{wu2019topological}. Correspondingly, 
the intralayer potential $V$ and interlayer tunneling strength $w$ is significantly larger than previously thought. 


\begin{table}[h]
{
\centering
\begin{tabular}{l|l|l|l|l|l|}
\hline
Materials & $\phi$ (deg) &	$V$ (meV) & $w$ (meV) & $m$ ($m_e$) & $a_0$ (Å) \\
\hline
$t$MoTe$_2$ & -91 & 11.2   & -13.3 & 0.62 & 3.52\\
$t$WSe$_2$ & 128 & 9 & -18 & 0.43 & 3.32 \\
\hline
\end{tabular}

\caption{Continuum model parameters extracted from density functional theory calculations. Parameters for $t$MoTe$_2$ are from this work and those for $t$WSe$_2$ are from \cite{devakul2021magic}.}\label{paramtable}
}
\end{table}

\begin{figure}
  \centering
\includegraphics[width=0.9\columnwidth]{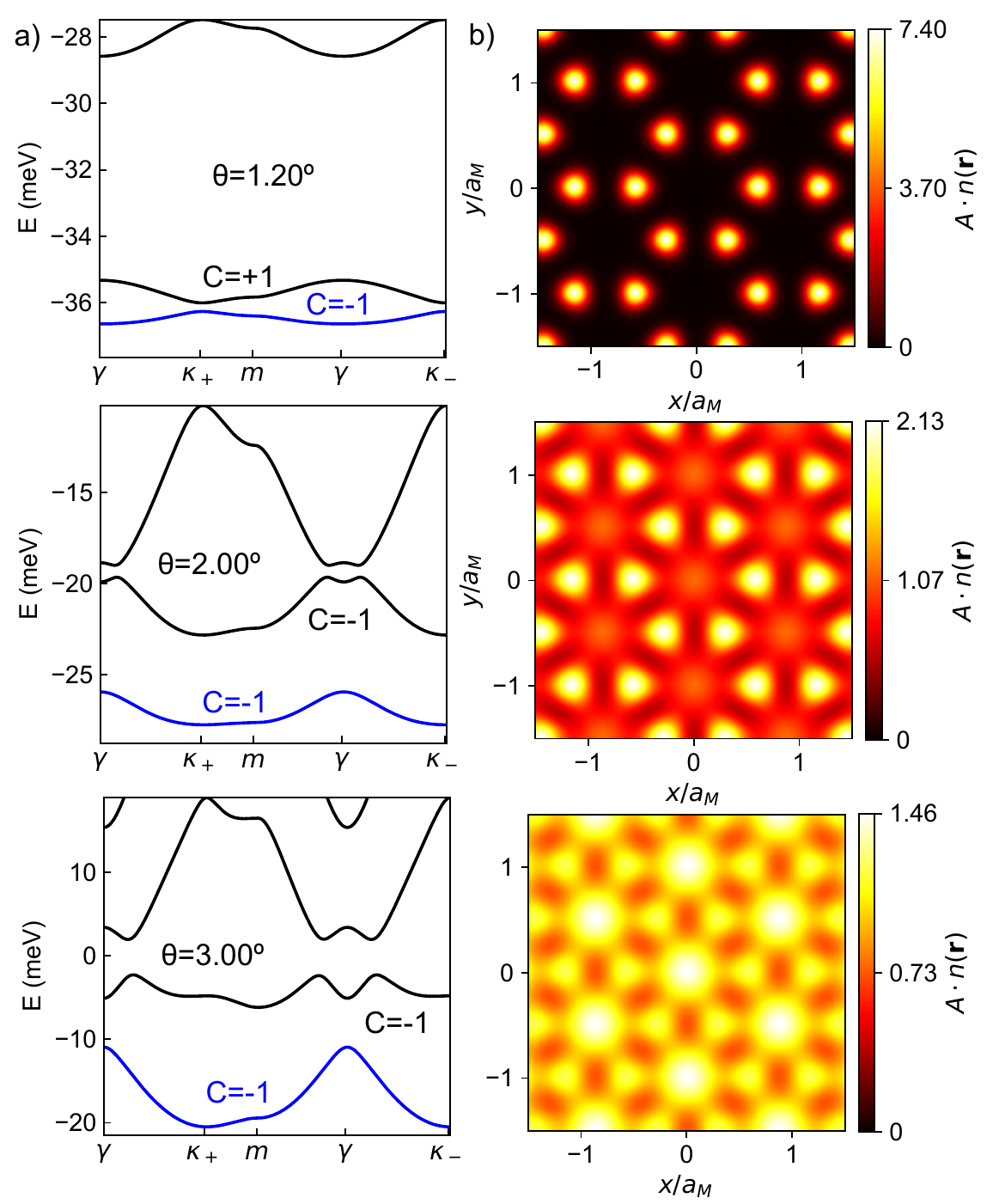}
  \caption{(a) Continuum model hole bands of $t$MoTe$_2$ in valley $K$ at several twist angles. Chern numbers of the first and second lowest bands are labeled in blue and black respectively. (b) Corresponding particle number density associated with the lowest band, A$n(\bm{r})=A\sum_{\bm{k}}|\psi_{1\bm{k}\uparrow}(\bm r)|^2$ where $A$ is the moiré unit cell area. The lowest band, onto which we project the continuum model Hilbert space in the exact diagonalization calculation, is highlighted in blue.} 
\label{fig:singleparticleintro}
\end{figure}


\emph{Twist angle dependence --} As we showed recently \cite{devakul2021magic}, the moir\'e band structure of twisted TMD homobilayers is highly tunable by the twist angle $\theta$, which controls the moiré period $a_M$ and thereby the ratio of kinetic to moiré potential energy. As the twist angle decreases, the moir\'e band structure evolves from nearly-free-electron-like to the tight-binding regime. In Fig. \ref{fig:singleparticleintro}, we show continuum model band structures and the corresponding charge densities at several twist angles. At $\theta=1.2^{\circ}$, the lowest two moiré bands have narrow widths $\sim 1$ meV and are well described by the Kane-Mele type tight binding model on the honeycomb lattice as we will elaborate later. The charge density of the lowest band exhibits sharp maxima at MX and XM stacking sites, which are local extrema of the intra-layer moiré potential and form a honeycomb lattice. We note that the layer polarization of these charge density peaks is strong and opposite between the two sublattices.

In the large angle limit where kinetic energy dominates, 
the charge density is more uniform. It exhibits shallower peaks on a triangular lattice of MM stacking regions where inter-layer tunneling is of maximum amplitude. The marked difference in the moir\'e band structures at low and high twist angles is evidenced by the lowest moiré band minimum changing from $\gamma$ to $\kappa_+ /\kappa_-$. As we showed recently \cite{devakul2021magic}, the crossover between the two regimes dictates the existence of a ``magic angle" at which the lowest moiré band becomes extremely flat.

\begin{figure}
  \centering
\includegraphics[width=\columnwidth]{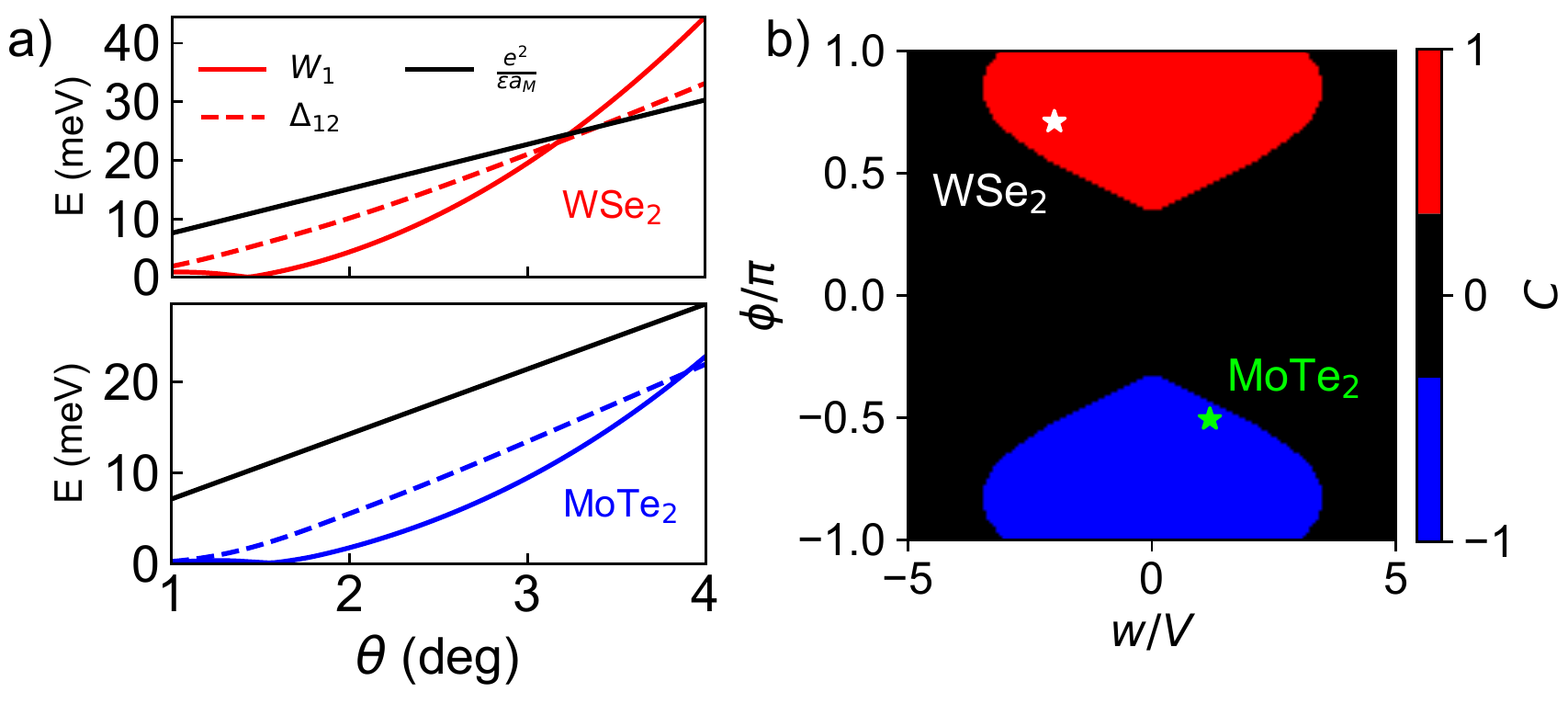}
  \caption{(a) Bandwidth of the lowest moiré band $(W_1)$ and difference in the average energy of two lowest moiré bands ($\Delta_{12}$) as a function of inter-layer twist angle $\theta$ for MoTe$_2$ and WSe$_2$. Also shown is a characteristic interaction energy scale $\frac{e^2}{\epsilon a_M}$ for $\epsilon=10$. (b) Chern number of the lowest moiré band as a function of the dimensionless ratio of the inter- and intra-layer moiré potential strengths $w/V$ as well as the phase parameter $\phi$ demonstrating that WSe$_2$ and MoTe$_2$ have opposite Chern numbers in a given valley.} 
\label{fig:bandenergyandchern}
\end{figure}

In Fig. \ref{fig:bandenergyandchern}, we also show the evolution of the width of the first band $W_1$ as well as the difference between the average energies of the first two bands $\Delta_{12}\equiv \sum_{\bm{k}}(\varepsilon_{2\bm{k}\uparrow}-\varepsilon_{1\bm{k}\uparrow})/\sum_{\bm{k}} 1$ as a function of twist-angle for both WSe$_2$ and MoTe$_2$. For angles $\gtrapprox 3^{\circ}$, the lowest moiré bands acquire significant dispersion $> 10$ meV. The bands of MoTe$_2$ are narrower than those of WSe$_2$. As we will later elaborate, as long as this bandwidth is small compared to the system's characteristic interaction energy, it plays an insignificant role in determining the many-body ground state. $\Delta_{12}$ in both cases monotonically increases with twist angle.

\emph{Band topology --} As first pointed out in the seminal work of Wu {\it et al} \cite{wu2019topological}, 
moiré bands of a given spin component in the continuum model for twisted bilayer TMD 
have finite Chern numbers that satisfy $C_\uparrow = -C_\downarrow \equiv C$ due to time reversal symmetry. 
As we will show later, the existence of topological bands in combination with small bandwidth at small twist angles is crucial for ferromagnetism and (integer and fractional) QAH states in this system. 

Remarkably, the topological character of moiré bands depends on the twist angle and shows two distinct regimes (see Fig. \ref{fig:singleparticleintro}). 
Previous theoretical studies have mostly focused on the ultra-small twist angle regime ($\theta<1.5^\circ$), where low-energy states are localized on a honeycomb lattice. Correspondingly, the lowest two moir\'e bands are isolated from higher bands and well described by a honeycomb lattice tight-binding model with Kane-Mele spin-orbit coupling \cite{kane2005z}. The first and second band of spin $\uparrow$ states in valley $K$ have Chern numbers 
$C_{\uparrow}=(-1,+1)$ respectively for MoTe$_2$ \cite{wu2019topological}. 


We now show that topological bands at larger twist angles have a different origin. 
This regime can be understood from a nearly-free-electron analysis. We treat the spatially varying intralayer moir\'e potential $V(\bm r)$ and interlayer tunneling $t(\bm r)$ as perturbations to the free particle gas. The leading effect of these perturbations is to 
induce superlattice gaps where free particle states with momenta $\bm k$ and $\bm k + \bm g$ are degenerate and coupled by $V(\bm{r})$, $t(\bm{r})$,   
leading to the formation of moir\'e bands. 
We calculate the superlattice gap and Bloch wavefunction of moir\'e bands at high symmetry points with a degenerate perturbation theory 
approach, first introduced in Ref.\cite{paul2023giant}, enabling us to determine the Chern number of moiré bands in a given valley in terms of the superlattice parameters $V, w, \phi$.

Fig. \ref{fig:bandenergyandchern} shows the Chern number $C$ thus obtained as a function of continuum model parameters $w/V$ and $\phi$ (without loss of generality $V$ is chosen to be positive). We further confirm by numerical calculation that the energy gap between the first and second moir\'e band remains finite over the entire range of twist angles from large to small, despite that the moir\'e band structure changes dramatically. Trivial bands ($C=0$) are found when the minima of the moir\'e potential $V(\bm r)$ are located at MM sites (which form a triangular lattice), whereas topological bands ($|C|=1$) are formed when (1) the intra-layer potential minima are located at MX/XM sites (which form a honeycomb lattice) and (2) the inter-layer tunneling $w$ is not too large compared to the intra-layer potential $V$. 

For a given valley/spin, the Chern numbers of the lowest moir\'e band at $\phi$ and $-\phi$ are opposite. Our large-scale DFT calculations find $\phi=-91^\circ$ for twisted bilayer MoTe$_2$ as presented above, and $\phi=128^\circ$ for twisted bilayer WSe$_2$ as shown in Ref.\cite{devakul2021magic}. Therefore, our theory predicts that the lowest moir\'e bands in twisted MoTe$_2$ and WSe$_2$ homobilayers have opposite Chern numbers in a given valley.

Our conclusion about band topology is further confirmed by examining the Bloch wavefunction of moir\'e bands in large-scale DFT calculation. The Chern number $\mod 3$ can be computed from the symmetry eigenvalues of spin-$\frac{1}{2}$ Bloch states under $C_{3z}$ rotation \cite{fang2012bulk}.
$C \mod 3 = \frac{3}{2\pi}\arg(- \lambda_{\kappa_+} \lambda_{\kappa_-} \lambda_{\gamma})$, 
where $\mathcal{R}_{\theta}$ is a rotation operator about the $z$-axis, and $\lambda_{\bm k} = \bra{\psi_{\bm k}}\mathcal{R}_{2\pi/3}\ket{\psi_{\bm k}}$ 
at three high symmetry points ${\bm k}=\kappa_+, \kappa_-, \gamma$ are $C_{3z}$ symmetry eigenvalues. 
The 
symmetry eigenvalues $\lambda$ for twisted bilayer MoTe$_2$ and WSe$_2$ are determined from DFT calculations (see Table ~\ref{table1}). 
The Chern number is then determined from 
the symmetry eigenvalues to be $C=-1$ and $1$ respectively.

\begin{table}[h]
{
\centering
\begin{tabular}{l|l|l|l|l|}
\hline
Materials &Band, Valley &	$\kappa_+$	& $\kappa_-$	& $\gamma$\\
\hline
$t$MoTe$_2$ &1, $K$   & $e^{i\pi/3}$	& $e^{i\pi/3}$ & $e^{-i\pi/3}$ \\
                  &1, $K^\prime$ & $e^{-i\pi/3}$	& $e^{-i\pi/3}$ & $e^{i\pi/3}$ \\
$t$WSe$_2$ &1, $K$   & $e^{i\pi/3}$	& $e^{i\pi/3}$ & $e^{i\pi}$ \\
 &1, $K^\prime$ & $e^{-i\pi/3}$ & $e^{-i\pi/3}$ & $e^{i\pi}$ \\
\hline
\end{tabular}

\caption{$C_{3z}$ eigenvalues of the topmost moiré bands from each valley, computed from large-scale DFT wavefunctions at high symmetry momentum points.}\label{table1}
}
\end{table}






Importantly, this difference in Chern number has observable consequences. According to the Streda formula, the Chern number determines the slope of the linear shift in carrier density with an applied magnetic field: $\frac{\partial n}{\partial B} =\frac{e}{hc}C$. We note that the measured sign of the $n$-$B$ slope in $t$MoTe$_2$ or $t$WSe$_2$ depends not only on the valley Chern number $C$, but also on the sign of the valley $g$-factor which determines the sign of the valley polarization under an applied $B$ field. Assuming that both  $t$MoTe$_2$ and $t$WSe$_2$ have valley g-factors of the same sign, 
our theory predicts that QAH states in the two systems will exhibit $n$-$B$ slopes of \emph{opposite} sign, see Fig. \ref{fig:intro}. 

Finally, we note that, as twist angle increases, the second band's Chern number changes sign from $-C$ to $C$ ($C$ is the Chern number of the first band) due to inversion with the third band at $\gamma$ (see Fig. \ref{fig:singleparticleintro}) \cite{wu2019topological, devakul2021magic}. This applies to both $t$MoTe$_2$ and $t$WSe$_2$.
Our DFT calculation shows that both the first and second bands of $t$MoTe$_2$ have the same Chern number in the twist angle range studied experimentally \cite{cai2023signatures}. As a result, we expect a double quantum spin Hall state at the filling of $n=4$ holes per unit cell \cite{devakul2021magic}, which can be experimentally detected through edge-state transport and current-induced edge spin polarization.

\section{Ising ferromagnetism and spin gap}

\begin{figure}
  \centering
\includegraphics[width=0.8\columnwidth]{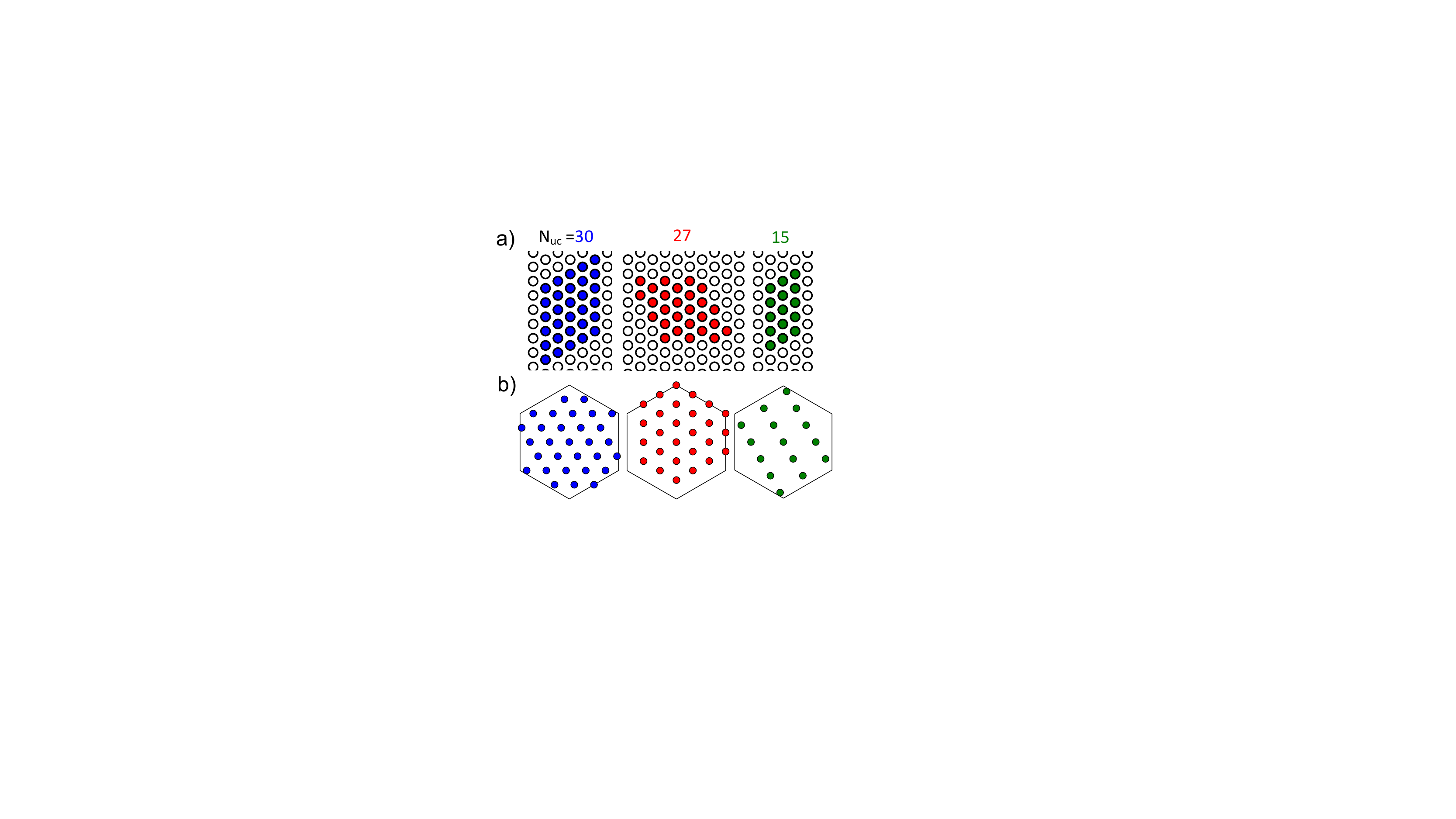}
  \caption{Finite sized clusters used in the exact diagonalization calculations in real space (a) and momentum space (b). The 27-unit-cell cluster can be viewed as a $9$-unit-cell cluster with a tripled unit cell.} 
\label{fig:clusters}
\end{figure}

We now turn to the many-body problem of a finite density of doped holes in twisted TMD homobilayers that interact with each other through Coulomb repulsion. 
The many-body continuum model Hamiltonian is given by
\begin{align}
\begin{split}
  H &= H_0 + V\\
  V &= \frac{1}{2}\sum_{\sigma,\sigma'} \int d {\bm r}d {\bm r}' \psi^\dag_\sigma(\bm r)\psi^\dag_{\sigma'}(\bm r')V(\bm r-\bm r')\psi_{\sigma'}(\bm r')\psi_\sigma(\bm r).
\end{split}
\end{align}

Here we use a long-range Coulomb interaction $V(\bm{r})=\frac{e^2}{\epsilon r}$ in contrast to previous studies of FQAH in TMD moiré superlattices using a dual-gate screened Coulomb interaction \cite{li2021spontaneous, crepel2022anomalous,morales2023pressure}. We perform an exact diagonalization calculation within the Fock space of the lowest Bloch band. 
Upon band projection, the many-body continuum model Hamiltonian is most conveniently written in momentum space as 

\begin{align}\label{eq:projham}
\begin{split}
    \tilde{H} 
    &= \sum_{\bm{k},\sigma}\varepsilon_{\bm{k}\sigma}c^{\dag}_{\bm{k}\sigma}c_{\bm{k}\sigma} \\
  &+ \frac{1}{2}\sum_{\bm{k}'\bm{p}'\bm{k}\bm{p},\sigma\sigma'}V_{\bm{k}'\bm{p}'\bm{k}\bm{p};\sigma\sigma'}c^{\dag}_{\bm{k}'\sigma}c^{\dag}_{\bm{p}'\sigma'}c_{\bm{p}\sigma'}c_{\bm{k}\sigma}
  \end{split}
\end{align}


where 
$c^{\dag}_{\bm{k}\sigma}$ creates a Bloch state in the lowest band at crystal momentum $\bm{k}$ and spin/valley $\sigma$ with corresponding single-particle energy $\varepsilon_{\bm{k}\sigma}$. $V_{\bm{k}'\bm{p}'\bm{k}\bm{p};\sigma\sigma'} \equiv \bra{\bm{k}'\sigma;\bm{p}'\sigma'}\hat{V}\ket{\bm{k} \sigma;\bm{p} \sigma'}$ are the corresponding matrix elements of the Coulomb interaction. Our choice to use the unscreened Coulomb interaction was motivated by simplicity and generality. We detail our methods in the supplement. 

Projection to the lowest band neglects band mixing  
and therefore is quantitatively accurate only when the ratio of the characteristic Coulomb energy $\frac{e^2}{\epsilon a_M}$ to the moiré band gap is small. However, band projection 
is known to be qualitatively correct in the study of fractional quantum Hall states in lowest Landau level, even when this dimensionless parameter is not small  
\cite{sodemann2013landau, sreejith2017surprising}. 
A follow up study of twisted TMD bilayers addressing band mixing 
is being prepared and will be presented elsewhere.  

In performing the calculation, we take advantage of the model's charge-$U(1)$, spin-$U(1)$, and translation symmetries to diagonalize within common eigenspaces of $N_h$, $S_z$, and crystal momentum. We 
use three clusters of different sizes and geometries illustrated in Fig. \ref{fig:clusters} with periodic boundary conditions. 

\begin{figure}
  \centering
\includegraphics[width=\columnwidth]{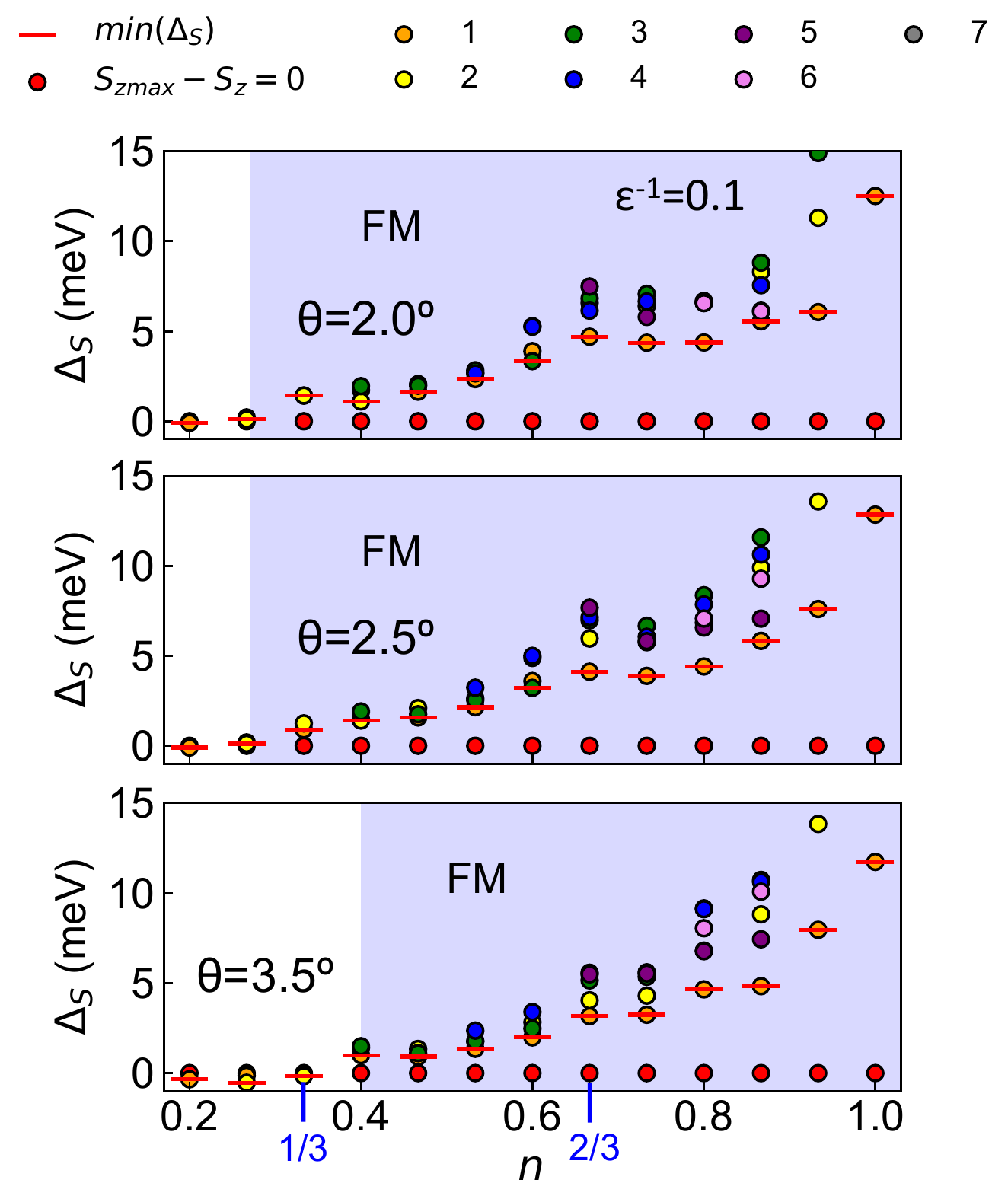}
  \caption{$\Delta_{S}\equiv E_{\text{min}}(S_z)-E_{\text{min}}(S_{z\text{max}})$ across all possible $S_z$ values within $15$ meV cutoff as a function of the filling factor $n\equiv N_h/N_{uc}$ on the 15-unit-cell cluster at fixed $\epsilon^{-1}=0.1$ and several $\theta$. min($\Delta_S$) is the minimum value of $\Delta_S$ for $S_z < S_{z\text{max}}$).} 
\label{fig:magnetism}
\end{figure}

We begin with an analysis of magnetism in $t$MoTe$_2$ across a broad range of filling factors $n\leq 1$. In Fig. \ref{fig:magnetism}, we plot $\Delta_{S}\equiv E_{\text{min}}(S_z)-E_{\text{min}}(S_{z\text{max}})$ for $S_z\geq 0$ as a function of the filling factor $n=N_h/N_{uc}$ on the 15-unit-cell torus with a fixed value of $\epsilon^{-1}=0.1$ and several twist angles. 
Here $E_{\text{min}}(S_z)$ is the minimum energy within a given $S_z$ sector, $N_h$ is the number of holes, and $N_{uc}$ is the number of moir\'e unit cells. 

At $\theta=2^\circ$, the lowest energy state is fully spin-polarized at all filling factors $0.27 \leq n \leq 1$ (the lower bound precision is limited by system size), showing robust spin/valley ferromagnetism in $t$MoTe$_2$. 
The spin gap (defined as the minimum of $\Delta_{S}$ with $S_z\neq S_{z\text{max}}$), which controls the Curie temperature and coercive field, 
is maximum at $n=1$ and generally decreases with decreasing $n$. Notably, we find the spin gap at $n=1$ exceeds 10 meV. 
A similar conclusion was reached for twisted TMD bilayers at smaller twist angles \cite{devakul2021magic, crepel2022anomalous}. 
Here, we find that Ising ferromagnetism and large spin gap ($>10$ meV) at $n=1$  persist to much larger twist angles, as shown $\theta=2.5^\circ$ and $3.5^\circ$. On the other hand, ferromagnetism at low filling $n<0.4$ is less robust and disappears at $\theta=3.5^\circ$ for $\epsilon^{-1}=0.1$.

At $\theta=2^\circ$, the spin gap clearly exhibits local maxima at filling factors $n=\frac{1}{3}$ and $\frac{2}{3}$, where FQAH states are formed as we will soon see. 
Notably, the spin gap at $n=\frac{2}{3}$ is much larger than at $n=\frac{1}{3}$. 
For $\epsilon^{-1}=0.1$, at $\theta=2.5^\circ$, the FQAH state only appears at $n=\frac{2}{3}$ where the spin gap is still weakly enhanced, but not at $n=\frac{1}{3}$. At $\theta=3.5^\circ$, the ground state at $n=\frac{2}{3}$ is a fully polarized Fermi liquid whose spin gap does not show any prominent feature, while the state at $n=\frac{1}{3}$ is non-magnetic.  

Consistent with our numerical findings, magnetic circular dichroism measurements on twisted bilayer MoTe$_2$ \cite{cai2023signatures} observed robust Ising ferromagnetism over a broad range of hole fillings between $n\sim 0.4$ and $1$, with a maximum Curie temperature of $15$ K at $n=1$. 
Moreover, the coercive field is enhanced at $n=\frac{2}{3}$. This agrees with the spin gap shown in Fig. \ref{fig:magnetism} and is associated with 
the formation of a FQAH state as we demonstrate below.  

Our calculation shows that Ising ferromagnetism in $t$MoTe$_2$ appears not only at $n=1, \frac{2}{3}$ and $\frac{1}{3}$, but throughout a broad range of filling factors below $n=1$ where the system is insulating or metallic alike. As a consequence of Ising ferromagnetism and Berry flux in moir\'e bands, we predict 
an anomalous Hall effect over a broad of fillings at and below $n=1$ (as found in $t$WSe$_2$ \cite{crepel2022anomalous}). 
In particular, quantized anomalous Hall effect is expected at $n=1$ and certain fractional filling factors that support FQAH insulators. 

From now on, we systematically study the many-body spectrum in the fully spin polarized sector at $n=\frac{2}{3}$ and $\frac{1}{3}$, for various twist angles and interaction strengths $\epsilon^{-1}$. We note that, at $n=\frac{1}{3}$ and large twist angles, the ground state may not be fully spin-polarized at zero field (see Fig. \ref{fig:magnetism}). We leave further investigation of spin physics at $n=\frac{1}{3}$ to future study.



\begin{figure}
  \centering
\includegraphics[width=\columnwidth]{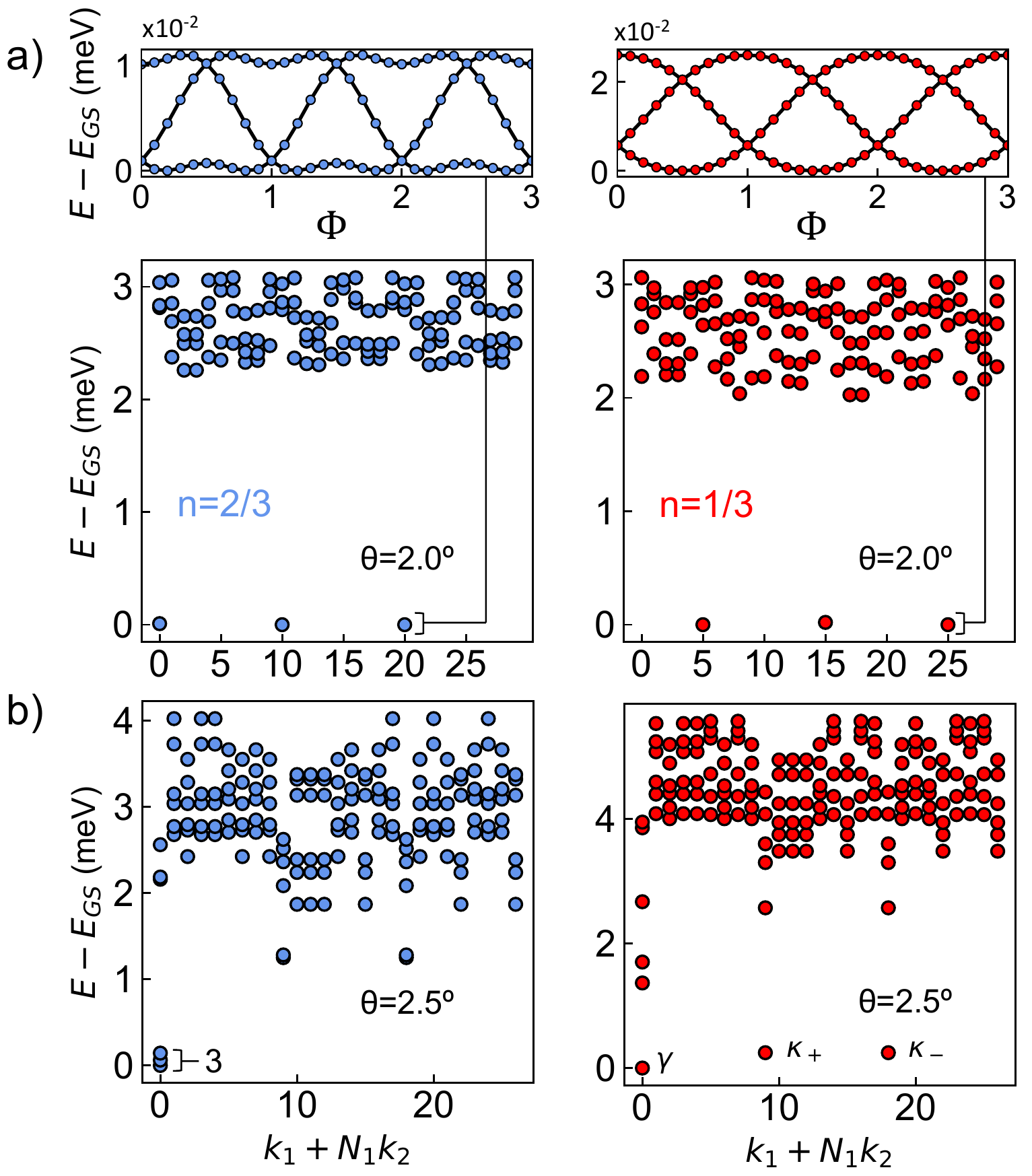}
  \caption{(a) Many-body spectra of $t$MoTe$_2$ within the fully polarized sector $(S_z = S_{\text{max}})$ on the 30-unit-cell cluster at $n=\frac{1}{3}$, $\frac{2}{3}$. We use $\theta=2.0^{\circ}$, $\epsilon=10$. At the top we show the ground state manifold's spectral flow under flux insertion demonstrating its FQAH nature. The 4 lowest states within each crystal momentum sector are shown. (b) Same as (a) except at a larger twist angle $\theta=2.5^{\circ}$ and on the 27-unit-cell cluster. The lowest 6 states within each momentum sector are shown. The spectrum at $n=\frac{2}{3}$ indicates FQAH whereas at $n=\frac{1}{3}$ indicates a CDW.}
\label{fig:mainED}
\end{figure}

\section{FQAH and competing phases}
In Fig. \ref{fig:mainED} (a), we show the many-body spectra obtained for $t$MoTe$_2$ on the 30-unit-cell cluster at $\theta=2^\circ$ and $n=\frac{1}{3}$, $\frac{2}{3}$ as a function of crystal momentum. We assign each crystal momentum $\bm{k}=k_1\bm T_1 + k_2 \bm T_2$ an integer index $k = k_1 + N_1k_2$ where $N_i$ is the number of crystal momenta along axis $i$. 
Here $\bm T_i = \frac{2\pi \epsilon_{ij}\bm L_j \times \hat{z}}{A}$ is a basis vector of crystal momentum, $\bm L_i$ defines the periodic boundary condition in real space, and $A=|\bm L_1 \times \bm L _2|$ is the system area. At both fillings, we find 3 nearly degenerate ground states separated by a sizable energy gap $\sim 2$ meV from excited states. The approximate ground state degeneracy matches the expected topological degeneracy 
of a fractional quantum Hall state on a torus. 
We note that imperfect ground state degeneracy is expected in a finite system. We have tested several cluster sizes with all other parameters fixed and find that the gap remains $\sim 2$ meV, indicating its presence in the thermodynamic limit. 
The many-body crystal momenta of the $n=\frac{1}{3}$, $\frac{2}{3}$ states -- having linear indices (5,15,25) and (0,10,20) respectively -- are in precise agreement with the generalized Pauli principle discussed in Ref. \cite{regnault2011fractional}.

In addition to the threefold ground state degeneracy, a necessary property of an $n=\frac{p}{q}$ fractional quantum Hall state is that its ground states on a torus permute upon insertion of one magnetic flux quantum 
such that each state returns to itself only after insertion of $q$ flux quanta. Flux insertion 
induces a shift in one component of the kinetic momentum $\bm \pi = \bm p + \hbar\Phi\bm T_{i}$ 
where $\Phi \equiv \frac{\phi}{\phi_0}$, $\phi$ is the inserted  flux, $\phi_0=\frac{h c}{e}$ is the flux quantum. 
In Fig. \ref{fig:mainED}(a), we show that both $n=\frac{1}{3}$ and $\frac{2}{3}$ exhibit this spectral flow, providing definitive evidence of their FQAH nature.

A change in the the twist angle $\theta$ 
induces a change in (1) the Bloch wavefunctions of the lowest band, (2) the band dispersion and width, 
and (3) the system's characteristic interaction energy scale $\frac{e^2}{\epsilon a_M}$. The band dispersion governs the kinetic energy $H_0$. At large twist angles where the lowest moir\'e band is highly dispersive, the ground state at fractional fillings is expected to be a Fermi liquid. 
The Bloch wavefunctions determine the form of the band-projected interaction $V$ through the Coulomb matrix elements. 
Therefore, 
a given filling factor, the system can exhibit distinct many-body ground states as a function of twist angle even when the band dispersion is neglected altogether. 
Thus, the influence of twist angle is multifold and needs 
systematic study. 


An obvious candidate ground state in the presence of strong, long-range Coulomb repulsion is a charge density wave (CDW). Such states are experimentally observed in TMD moiré hetero-bilayers where they are known as generalized Wigner crystals \cite{regan2020mott, xu2020correlated, li2021imaging, huang2021correlated}. To address the possible competition between FQAH and CDW with exact diagonalization, it is essential to choose a cluster that accommodates a tripled unit cell or, equivalently, samples $\gamma, \kappa_+$, and $\kappa_-$. The 27-unit-cell cluster depicted in Fig. \ref{fig:clusters} satisfies this criterion. In Fig. \ref{fig:mainED}(b) we show spectra obtained at a larger twist angle $\theta=2.5^\circ$ using this cluster. At $n=\frac{2}{3}$, we find three nearly degenerate ground states at $\gamma$, indicative of FQAH. On the other hand, at $n = \frac{1}{3}$, we find three nearly degenerate states with one at each of $\gamma, \kappa_+$ and $\kappa_-$ (the center and corners of the moiré Brillouin zone, respectively). These are the momenta appropriate to a charge density wave with a tripled unit cell because they fold back to $\gamma$ in the symmetry-broken Brillouin zone. 

\begin{figure}
  \centering
\includegraphics[width=0.8\columnwidth]{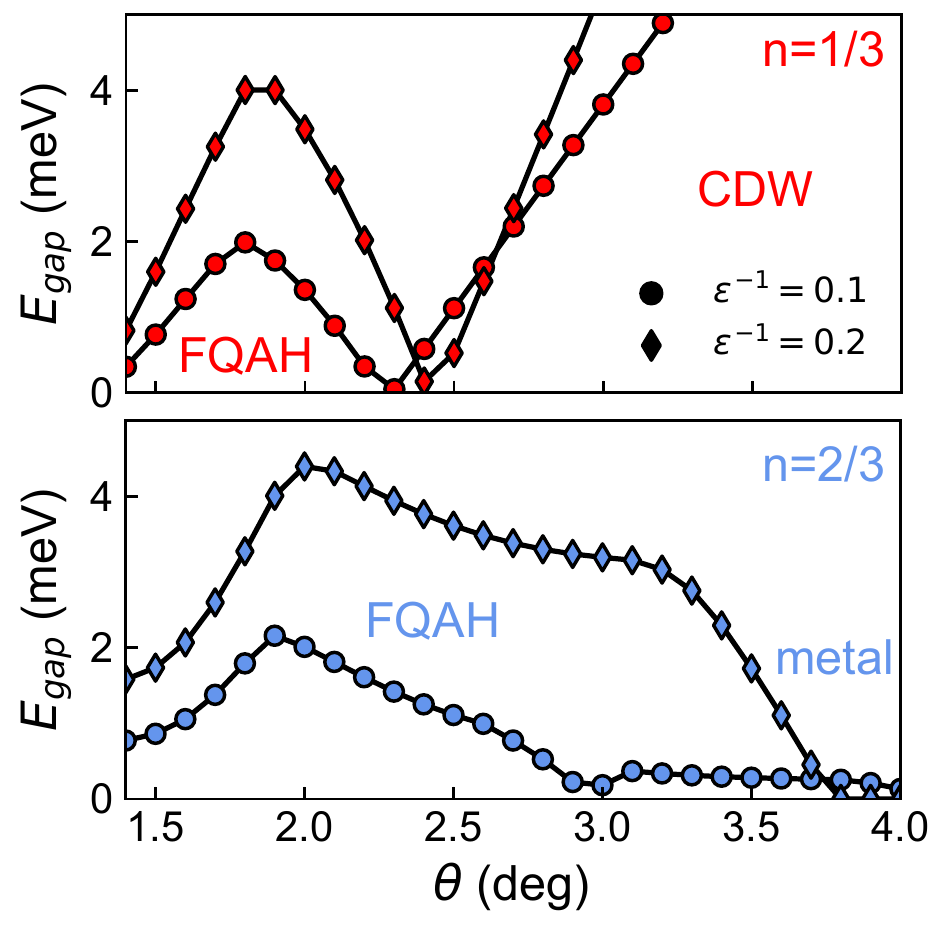}
  \caption{$E_{\text{gap}} \equiv E_4-E_3$ where $E_i$ is the fully spin-polarized state of $i^{th}$ lowest energy as a function of twist angle $\theta$ for $n=\frac{1}{3},\, \frac{2}{3}$ and two values of $\epsilon^{-1}=0.1, \,0.2$.} 
\label{fig:thetasweep}
\end{figure}

To reveal the influence of twist angle on many-body ground states at $n=\frac{1}{3}$ and $\frac{2}{3}$, 
in Fig. \ref{fig:thetasweep} we plot the energy gap $E_{\text{gap}}=E_4-E_3$ as a function of $\theta$, where $E_i$ is the $i^{th}$ lowest energy with a fixed $N_h$ and maximum spin $S_z$. Two values of the dielectric constant, $\epsilon^{-1}=0.1,\,0.2$ are used. 
When the system is a correlated insulator with threefold ground state degeneracy such as FQAH or CDW, $E_{\text{gap}}$ is indicative of its robustness. 

For $\epsilon^{-1}=0.1$, we see that both the $n=\frac{1}{3},\, \frac{2}{3}$ states exhibit maxima in $E_{\text{gap}}$ near $\theta=1.8^{\circ}$. 
Beyond $\theta \approx 1.8^{\circ}$, 
$E_{\text{gap}}$ decreases at both fractions, but more rapidly so at $n=\frac{1}{3}$ where it reaches zero near $\theta \approx 2.3^{\circ}$ and then increases again. The many-body spectra on both sides of this gap-closing transition (not shown) have 
three nearly degenerate ground states. However, the ground states at $\theta<2.3^\circ$ have the crystal momenta of the FQAH state as shown in Fig. \ref{fig:mainED}(a), whereas those at $\theta>2.3^\circ$ have the crystal momenta of the CDW state as shown in Fig. \ref{fig:mainED}(b). Thus, we conclude that at the fractional filling $n=\frac{1}{3}$, a quantum phase transition between FQAH and CDW occurs around $\theta \approx 2.3^{\circ}$. 

The situation is markedly different at $n=\frac{2}{3}$. In this case, $E_{\text{gap}}$ remains finite until $\theta\approx 3.0^{\circ}$, beyond which point it is 
small. The many-body spectrum shows a continuum of states at low energy, indicating a metallic phase in the thermodynamic limit. 
These results 
clearly show that the FQAH state at $n=\frac{2}{3}$, previously overlooked in theoretical studies of twisted homobilayer TMD \cite{li2021spontaneous,crepel2022anomalous}, persists to a substantially higher twist angle than that at $n=\frac{1}{3}$.

When $\epsilon^{-1}=0.2$, the dependence of the $n=\frac{1}{3}$ state on $\theta$ is largely similar to when $\epsilon^{-1}=0.1$, save for an expected increase of  $E_{\text{gap}}$ due to the increased Coulomb interaction.
On the other hand, at 
$n=\frac{2}{3}$, the increased interaction 
pushes the FQAH-metal transition 
to $\theta \approx 3.8^\circ$, thereby significantly expanding the twist angle range of the FQAH state.  

These numerical results provide valuable insight into the competition between FQAH, CDW, and metallic phases. At small twist angles, 
the bands are narrow enough (see Fig. \ref{fig:bandenergyandchern}) that for both $\epsilon^{-1}=0.1$ and $0.2$ the system is in its flat band limit $\frac{e^2}{\epsilon a_M}/W \gg 1$. The many-body ground state is thus determined primarily by the projected interaction term which is in turn determined by the Bloch wavefunctions. For $\theta \lessapprox 2.3^{\circ}$, FQAH state is preferred by at both fillings. 
On the other hand, at large twist angles, the bandwidth becomes sizable and is crucial in the competition between FQAH and metallic phase at $n=\frac{2}{3}$.  

\begin{figure}
  \centering
\includegraphics[width=\columnwidth]{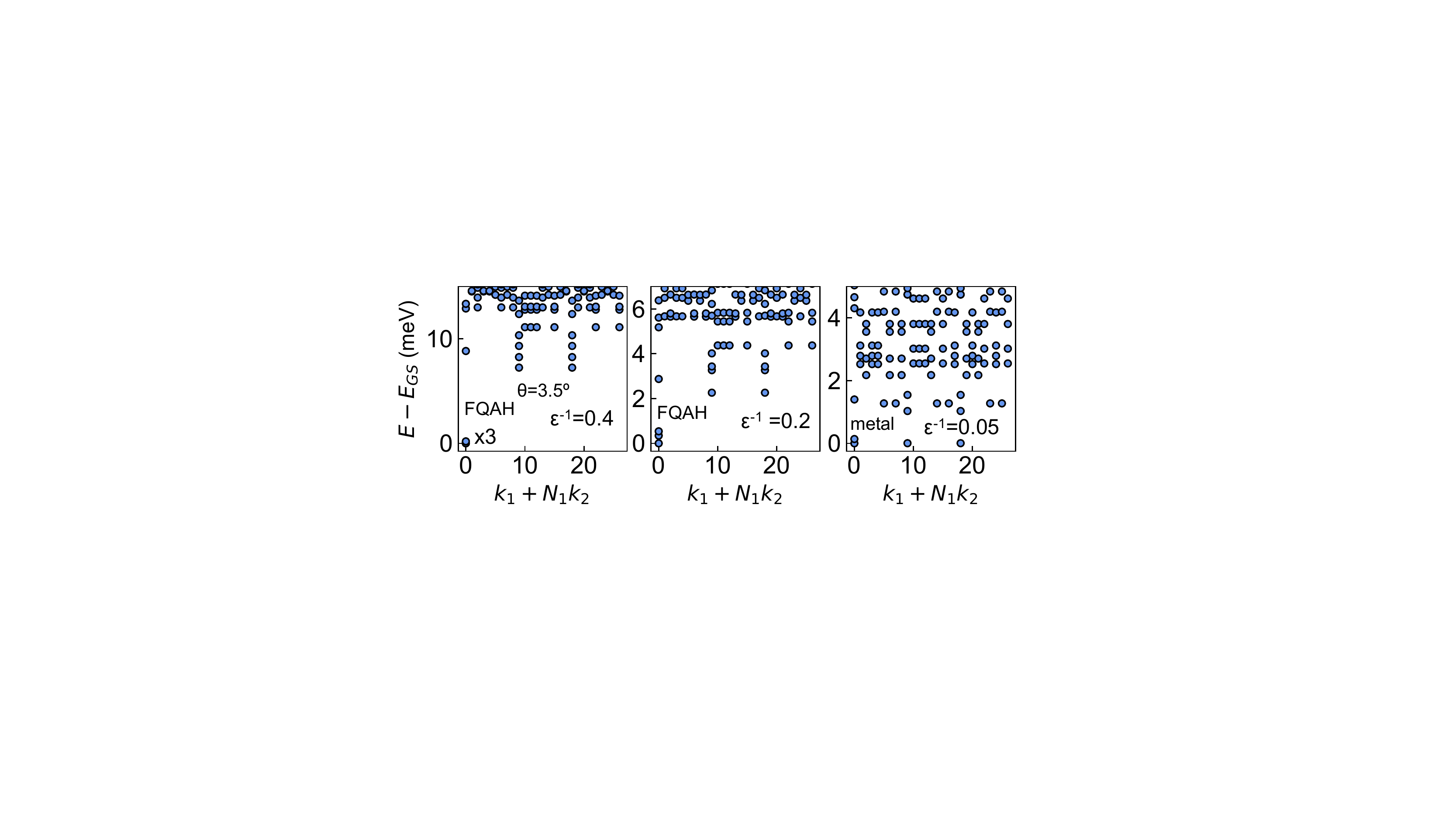}
  \caption{Low-lying spectra at $\theta=3.5^{\circ}$ for several values of the inverse dielectric constant $\epsilon^{-1}$ at $n=\frac{2}{3}$. For small $\epsilon^{-1}=0.1$ (weak interaction) the system is not in a FQAH state whereas for $\epsilon^{-1} \gtrapprox 0.2$, it is. All energy levels in the fully spin-polarized sector and the given window are shown.} 
\label{fig:epssweep}
\end{figure}

To disentangle the effect of bandwidth from that of Bloch wavefunction, we study the FQAH-metal transition at $n=\frac{2}{3}$ for a fixed $\theta=3.5^\circ$, tuned by the interaction strength $\epsilon^{-1}$. 
Changing $\epsilon^{-1}$ does not affect the band wavefunction, but tunes the ratio of bandwidth and interaction energy. 
Fig. \ref{fig:epssweep} shows the many-body spectra at $\epsilon^{-1}=0.4, 0.2, 0.05$. 
While $\epsilon^{-1}=0.4$ is likely larger than experimental values, 
it provides useful insight into the strong coupling limit 
in a similar spirit to the band-flattening approach 
\cite{neupert2011fractional}.

Starting from the strong interaction limit $\epsilon^{-1}=0.4$, it is clear that $n=\frac{2}{3}$ exhibits FQAH with $3$ well isolated, nearly degenerate states at $\gamma$ as expected from the generalized Pauli principle rules applied to this cluster geometry. 
As the interaction decreases, the energy gap at momenta $\kappa_+$ and $\kappa_-$ softens ($\epsilon^{-1}=0.2$), before the metallic state with a continuum of low-lying states appears ($\epsilon^{-1}=0.05$). 
The nature of this FQAH-metal transition at $n=\frac{2}{3}$ is an interesting and important question that calls for further study. 

\section{
$n=\frac{1}{3}$ versus $\frac{2}{3}$
}

We have shown by exact diagonalization study that AA-stacked TMD moiré homobilayers exhibit robust Ising spin/valley ferromagnetism across a wide range of carrier densities within the lowest moiré band. 
Since the valley-polarized moiré bands carry  finite Chern number, anomalous Hall effect is expected throughout. At particular fractional filling factors $n=\frac{p}{q}=\frac{1}{3}$,\ $\frac{2}{3}$ we predict fractional quantum anomalous Hall states with corresponding quantized Hall conductances $\sigma_H = \frac{p}{q}\frac{e^2}{h}$. Using continuum model parameters obtained from our first-principles band structure calculations, our study finds that the topological gap of FQAH states in twisted bilayer MoTe$_2$ is largest near $\theta\approx 2^{\circ}$.

At larger twist angles, the $n=\frac{1}{3}$ state gives way to a CDW near $\theta \approx 2.3^{\circ}$ whereas the $n=\frac{2}{3}$ state persists to a larger angle, eventually becoming metallic. As interaction strength increases, the FQAH regime at $n=\frac{2}{3}$ extends to higher angles, suggesting that the FQAH-metal transition is primarily bandwidth controlled. On the other hand, the critical angle of the FQAH-CDW transition at $n=\frac{1}{3}$ is weakly dependent on interaction strength, indicating that it is instead controlled by a change in the Bloch wavefunctions.  

The difference between $n=\frac{1}{3}$ and $\frac{2}{3}$ filling states is noteworthy and interesting. 
Recall that the ground states of a Landau level at filling factors $n$ and $1-n$ are simply related by a particle-hole transformation that leaves the projected Hamiltonian invariant. 
 This is not the case in our system. In particular, at large twist angles, the $\frac{1}{3}$- and $\frac{2}{3}$-filling ground states are distinct phases of matter: CDW and Fermi liquid respectively. 

To understand the contrast between $n=\frac{1}{3}$ and $\frac{2}{3}$, we note that the band-projected Hamiltonian within the fully spin-polarized sector $S_z=S_{z\text{max}}$ is {\it not} symmetric under particle-hole transformation $c_\uparrow^\dagger(\bm{r}) \rightarrow d_\uparrow(\bm{r})$. 
In particular, 
unlike a single particle added to an otherwise empty moiré band, a single particle removed from an otherwise full moiré band has an interaction-induced dispersion present even when the bare bandwidth $W$ vanishes \cite{guinea2018electrostatic,abouelkomsan2020particle,kang2021cascades}.


We now show that the interaction-induced asymmetry between particle and hole dispersion provides a natural explanation of the difference between $n=\frac{1}{3}$ and $\frac{2}{3}$ filling states found in our exact diagonalization study. 
Notably, we find that at large twist angles, a single hole at $n=1$ is more dispersive than a single particle at $n=0$. Therefore, in the presence of Coulomb interaction, the system at low filling $\delta\ll 1$ is more susceptible to Wigner crystallization into a CDW state than at the filling $1-\delta$. This explains our finding of CDW at $n=\frac{1}{3}$ and Fermi liquid at $\frac{2}{3}$.

\section{Discussion}
We have also obtained strong numerical evidence for FQAH states at filling factors $n=\frac{2}{5}, \frac{3}{5}$. 
For $\theta=2^{\circ}$, $\epsilon^{-1}=0.1$ and at both fillings, calculations on the 30-unit cell-cluster show fivefold nearly degenerate ground states 
separated from the continuum by an energy gap $> 1$ meV. 

As noted above, previous exact diagonalization studies of FQAH states in AA-stacked TMD homo-bilayers have focused on the ultra-low twist-angle regime $\theta < 1.5^{\circ}$ \cite{li2021spontaneous, crepel2022anomalous, morales2023pressure}. This is where the lowest moir\'e band satisfies various conditions purported in the literature to support FQAH states, including nearly vanishing bandwidth and quantum geometric properties---Berry curvature uniformity and ``trace condition" 
\cite{roy2014band, claassen2015position, wang2021exact}. 


In comparison with previous studies, our work explores larger twist angles and additional filling fraction $n=\frac{2}{3}$ where the FQAH state has recently been observed. In addition, we elucidate the origin of band topology at large twist angle through degenerate perturbation theory analysis.
Our results clearly show that FQAH states in twisted bilayer MoTe$_2$ extend to significantly larger twist angles where 1) the lowest moir\'e band 
has significant dispersion and 2) the first and second moiré bands within a given valley have the same Chern number contrary to the Kane-Mele regime at small twist angle. Nonetheless, the Coulomb energy scale  $e^2/(\epsilon a_M)$ suffices to support CDW and FQAH phases. 
 Remarkably, at larger twist angles, FQAH state is found at $\frac{2}{3}$ filling, whereas CDW is found at $\frac{1}{3}$. In addition, we find that FQAH states are accompanied by spin gap enhancements, in agreement with the experimental observations of Ref. \cite{cai2023signatures}. 
Our findings point to the surprising richness and robustness of FQAH physics beyond flat band and ideal quantum geometry.

  
Consistent with recent experimental observations \cite{cai2023signatures, foutty2023mapping}, our band-projected exact diagonalization study also shows a robust integer quantum anomalous Hall effect at $n=1$ protected by a large spin gap.
Here, we also note the possibility of topologically trivial, layer-polarized states at $n=1$ as well as fractional fillings, especially at small twist angles where the band gap is small. To faithfully describe such states requires the inclusion of at least two lowest bands \cite{abouelkomsan2022multiferroicity}, which goes beyond our single-band calculation. As mentioned above, an investigation of band mixing effects is currently underway and results will be presented elsewhere. 

  A straightforward extension of our work is to understand the influence of displacement field on the QAH states. Generally speaking, stronger displacement field should drive the system into topologically trivial, layer-polarized states \cite{wu2019topological}. Very recently, the ability to tune a topological phase transition from the integer QAH state to a topologically trivial state at $n=1$ has been demonstrated experimentally in twisted bilayer WSe$_2$ \cite{foutty2023mapping}. 

  Finally, we highlight the prospect of QAH beyond the lowest moiré band. Indeed, integer QAH at $n=3$ has already been observed in $t$WSe$_2$ \cite{foutty2023mapping}, and the possibility of $n\geq 1$ fractional states is enticing.


\section{Acknowledgement}
We thank Xiaodong Xu for collaboration on a closely related experiment on $t$MoTe$_2$ \cite{cai2023signatures}, Ben Foutty and Ben Feldman for collaboration on a closely related experiment on $t$WSe$_2$ \cite{foutty2023mapping}, Valentin Crépel for previous collaboration on related theoretical works \cite{devakul2021magic,crepel2022anomalous}, as well as 
Di Luo and Patrick Ledwith for helpful discussions. This work was supported by the Air Force Office of Scientific Research (AFOSR) under award FA9550-22-1-0432 and the David and Lucile Packard Foundation. Y.Z. acknowledges support from the start-up fund at the University of Tennessee. F. A acknowledges support from the KAUST Gifted Students Program and the Undergraduate Research Opportunities Program at MIT. 

{\it Note added}: we recently became aware of independent work on similar topics \cite{wang23toappear}. Also, a related experimental development on integer and fractional quantum anomalous Hall states in $t$MoTe$_2$ has been reported \cite{zeng2023integer}.


%

\end{document}


\title{Supplementary material for \emph{Fractional quantum anomalous Hall states in twisted  
bilayer MoTe$_2$ and WSe$_2$}}
\author{Aidan P. Reddy}
\email{areddy@mit.edu}
\affiliation{Department of Physics, Massachusetts Institute of Technology, Cambridge, Massachusetts 02139, USA}
\author{Faisal Alsallom}
\affiliation{Department of Physics, Massachusetts Institute of Technology, Cambridge, Massachusetts 02139, USA}
\author{Yang Zhang}
\affiliation{Department of Physics and Astronomy, University of Tennessee, Knoxville, Tennessee 37996, USA}
\affiliation{Min H. Kao Department of Electrical Engineering and Computer Science, University of Tennessee, Knoxville, Tennessee 37996, USA}
\author{Trithep Devakul}
\affiliation{Department of Physics, Massachusetts Institute of Technology, Cambridge, Massachusetts 02139, USA}
\author{Liang Fu}
\email{liangfu@mit.edu}
\affiliation{Department of Physics, Massachusetts Institute of Technology, Cambridge, Massachusetts 02139, USA}
\date{\today}
\date{\today}
\maketitle
\tableofcontents

\section{Density functional theory calculation and continuum model}
We study TMD homobilayers with a small twist angle starting from AA stacking, where every metal (M) or chalcogen (X) atom  on  the top layer is aligned with the same type of atom on the bottom layer. 
Within a local region of a twisted bilayer, the atom configuration is identical to that of an untwisted bilayer, where one layer is laterally shifted relative to the other layer by a corresponding displacement vector ${\bm d}_0$. For this reason, the moir\'e band structures of twisted TMD bilayers can be constructed from a family of untwisted bilayers at various ${\bm d}_0$, all having $1\times 1$ unit cell. Our analysis thus starts from untwisted bilayers.

In particular, ${\bm d}_0=0, -\left({\bm a}_{1}+{\bm a}_{2}\right) /3, \left({\bm a}_{1}+{\bm a}_{2}\right) /3$, where ${\bm a}_{1,2}$ is the primitive lattice vector  for untwisted bilayers, correspond to three high-symmetry
stacking configurations of untwisted TMD bilayers, which we refer to as MM, XM, MX. In MM (MX) stacking, the M atom on the top layer is locally aligned with the M (X) atom on the bottom layer, likewise for XM. The bilayer structure in these stacking configurations is invariant under three-fold rotation around the $z$ axis.

Density functional calculations are performed using generalized gradient approximation with SCAN density functional ~\cite{peng2016versatile} with dDsC dispersion correction method, as implemented in the Vienna Ab initio Simulation Package. Pseudopotentials are used to describe the electron-ion interactions. We first construct the zero-twisted angle MoTe$_2$/MoTe$_2$ bilayer at MM and MX lateral configurations with vacuum spacing larger than 20\AA{} to avoid artificial interaction between the periodic images along the z direction. Lattice constant 3.52\AA{} is taken from bulk structures. The structure relaxation is performed with force on each atom less than 0.001 eV/A. We use $12\times 12 \times 1$ for structure relaxation and self-consistent calculation. The more accurate SCAN+ dDsC dispersion correction method gives the relaxed layer distances as 7.02\AA{} (close to the bulk layer distance 6.98\AA{} in 2H structures) and 7.80\AA{} for MX and MM stacking structures, respectively. 

\begin{figure}
\includegraphics[width=0.5\columnwidth]{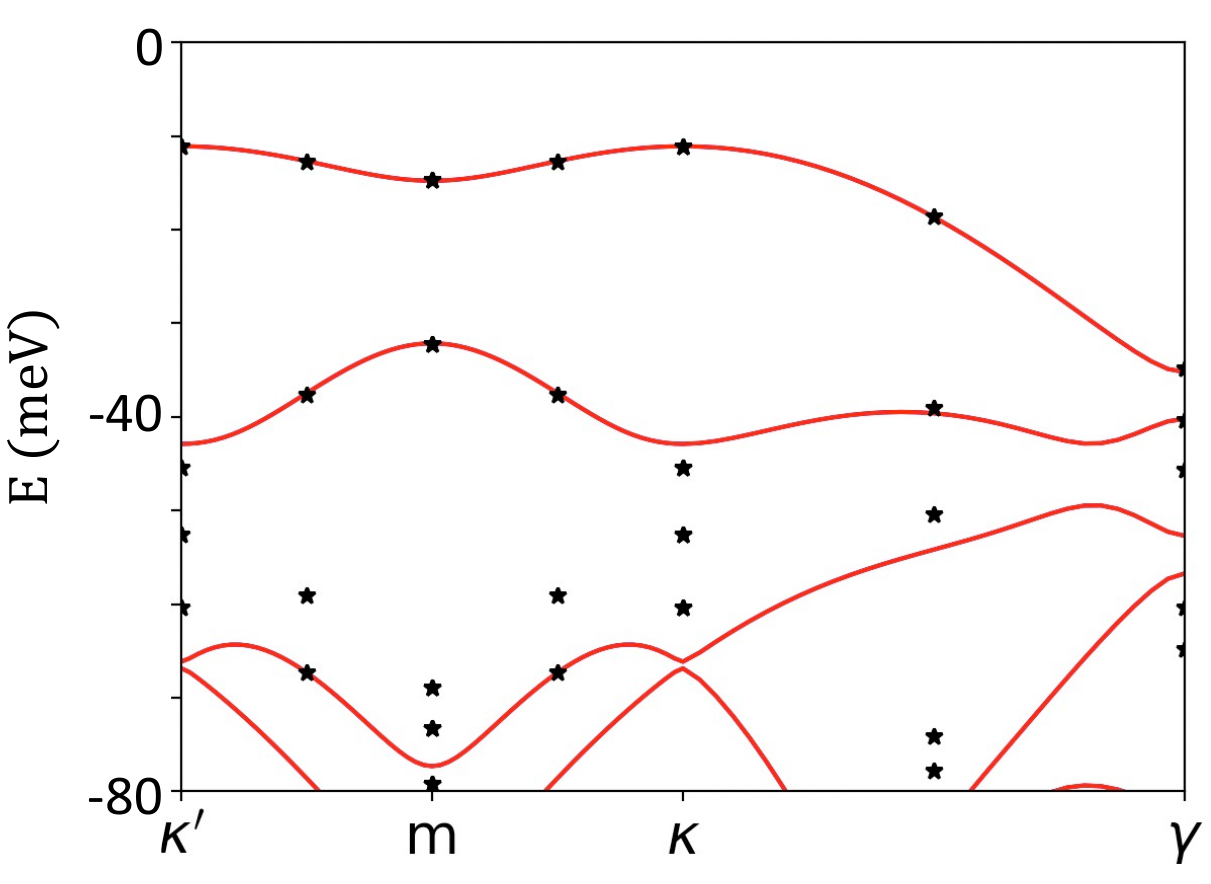}
\caption{The continuum band structure (red lines) is plotted in comparison with large scale DFT calculations (black dots) at twist angle $\theta=3.89^\circ$.}
\label{figS1}
\end{figure}

To compare with moir\'e band structure, we further construct the commensurate structures with twist angle $4.4^\circ$ with 1014 atoms, and perform large scale DFT calculation to relax the moir\'e superlattice. The spacial dependent layer distance profile has been discussed in the main text, and fits nicely with the untwisted MM and MX stacking structures. The moir\'e band structure with spin orbital coupling agrees remarkably well with the continuum model derived from untwisted calculations, and the effective mass is chosen as 0.62$m_e$. 

For the experimental twist angle $3.89^\circ$ with 1302 atoms, we apply the less expensive GGA functional with van der Waals correlation as Tkatchenko-Scheffler method with iterative Hirshfeld partitioning. The fitted continuum model band structure and DFT band structure are shown in Fig. \ref{figS1}. And the corresponding parameters are $V=7.5$meV, $\phi=-100^\circ$ and $w=-11.3$meV, with the bulk lattice constant $a_0 = 3.52\text{\AA{}}$ and the effective mass $m^* = 0.62 m_e$. While the top two moir\'e valence bands agrees very well with the continuum model, the rest of valence bands have large deviations. This indicates that there are additional moir\'e bands from $\Gamma$ valley and that higher harmonics should be included when considering third moir\'e valence band from $K$ valley. We will study the variation of continuum model parameter under different van der Waals correlations and lower moir\'e valence bands in future work.

\section{Continuum model band-projected exact diagonalization}
\subsection{Setup}
We numerically diagonalize the continuum model Hamiltonian projected to the lowest moiré band with a Coulomb interaction $V(\bm{r}) = \frac{e^2}{\epsilon |\bm{r}|}$. The band-projected Hamiltonian is
\begin{align}\label{eq:projham}
\begin{split}
    \tilde{H} &= \sum_{\bm{k},\sigma}\varepsilon_{\bm{k}\sigma}c^{\dag}_{\bm{k}\sigma}c_{\bm{k}\sigma} + \frac{1}{2}\sum_{\bm{k}'\bm{p}'\bm{k}\bm{p},\sigma\sigma'}V_{\bm{k}'\bm{p}'\bm{k}\bm{p};\sigma\sigma'}c^{\dag}_{\bm{k}'\sigma}c^{\dag}_{\bm{p}'\sigma'}c_{\bm{p}\sigma'}c_{\bm{k}\sigma}
  \end{split}
\end{align}
where $c^{\dag}_{\bm{k}\sigma}$ creates a hole with moiré crystal momentum $\bm{k}$ and spin/valley quantum number $\sigma$ with band energy $\varepsilon_{\bm{k}\sigma}$. $V_{\bm{k}'\bm{p}'\bm{k}\bm{p};\sigma\sigma'} \equiv \bra{\bm{k}'\sigma;\bm{p}'\sigma'}\hat{V}\ket{\bm{k} \sigma;\bm{p} \sigma'}$ are the Bloch state interaction matrix elements. Band-projected ``exact" diagonalization is a variational method for approximating low-energy many-body states. The Fock space of the lowest moiré band is its variational subspace. It is designed to be exact in the absence of interactions but will be inexact in their presence where the exact low-lying states will generally have nonzero weight outside of the lowest band Fock space. It is an accurate approximation to the extent that the true low-lying many-body states lie in the lowest band Fock space.

Our first computational step is to diagonalize the single-particle continuum model Hamiltonian (Eq. 2 of the main text) in a basis of plane wave states labeled by total momentum measured relative to the moiré Brillouin zone $\gamma$ point written as a sum of moiré crystal $\bm{k}$ and reciprocal lattice $\bm{g}$ momenta, layer, and $S_z$ spin = valley $\ket{\bm{k}+\bm{g},l,\sigma}$. (These ``plane wave" states really represent Bloch waves of the isolated monolayers and their ``total" momentum is really monolayer crystal momentum measured relative to the moiré Brillouin zone $\gamma$ point.) Here we provide an explicit equation for $\mathcal{H}_{\downarrow}$.
\begin{align}\label{eq:CMHam}
  \mathcal{H}_{\downarrow} = \begin{pmatrix}
  \frac{\hbar^2(- i \nabla + \kappa_+)^2}{2m} + V_1(\bm{r}) & t^{\dag}(\bm{r}) \\
  t(\bm{r}) & \frac{\hbar^2( - i\nabla + \kappa_-)^2}{2m} + V_2(\bm{r})
  \end{pmatrix}
\end{align}
Equations for all terms therein, $\mathcal{H}_{\uparrow}$, $H_0$ are in the main text. Given $\mathcal{H_{\uparrow}}$, $\mathcal{H}_\downarrow$ can be derived by enforcing that $\mathcal{T}H_0\mathcal{T}^{-1}=H_0$ where $\mathcal{T}$ is the time reversal operator (see, for instance, chapter 4 of Ref. \cite{bernevig2013topological}). In this basis, the matrix elements of the continuum model Hamiltonian are
\begin{align}
    \begin{split}
    \bra{\bm{k}+\bm{g}',l',\uparrow}H_0\ket{\bm{k}+\bm{g},l,\uparrow} &= \delta_{l'l}\left(\delta_{\bm{g}'\bm{g}}\frac{\hbar^2(\bm{k} - \kappa_l)^2}{2m} + V^{l}_{\bm{g}'\bm{g}}\right) + w t^{l'l}_{\bm{g}'\bm{g}},
    \end{split}
\end{align}
\begin{align}
    \begin{split}
    \bra{\bm{k}+\bm{g}',l',\downarrow}H_0\ket{\bm{k}+\bm{g},l,\downarrow} &= \delta_{l'l}\left(\delta_{\bm{g}'\bm{g}}\frac{\hbar^2(\bm{k} + \kappa_l)^2}{2m} + V^{l}_{\bm{g}'\bm{g}}\right) + w t^{ll'}_{\bm{g}'\bm{g}}
    \end{split}
\end{align}
where
\begin{align}
    V^{l}_{\bm{g}'\bm{g}} = -V\sum_{j=1}^{6}\delta_{(\bm{g}'-\bm{g})\bm{g}_j}e^{i(-1)^{(j+1)}\phi_{l}},
\end{align}
\begin{align}
    t^{l'l}_{\bm{g'}\bm{g}} = w\left[\delta_{l'1}\delta_{l2}\left(\delta_{\bm{g}'\bm{g}} + \delta_{\bm{g}'(\bm{g}+\bm{g}_2)}+\delta_{\bm{g}'(\bm{g}+\bm{g}_3)}\right)  + \delta_{l'2}\delta_{l1}\left(\delta_{\bm{g}'\bm{g}} + \delta_{\bm{g}'(\bm{g}-\bm{g}_2)}+\delta_{\bm{g}'(\bm{g}-\bm{g}_3)}\right)\right].
\end{align}
Here $\bm{g}_i=\frac{4\pi}{\sqrt{3}a_M}(\cos\frac{\pi(i-1)}{3},\sin\frac{\pi(i-1)}{3})$, $\kappa_- = \frac{\bm g_1 +\bm g_6}{3}$, $\kappa_+ = \frac{\bm g_1 +\bm g_2}{3}$, $\kappa_1=\kappa_+$, $\kappa_2=\kappa_-$, and $\phi_2=-\phi_1=\phi$. All matrix elements that are off-diagonal in crystal momentum or spin/valley vanish.
\par Numerically diagonalizing this single-particle Hamiltonian yields Bloch states labeled by moiré crystal momentum and spin=valley quantum numbers:
\begin{equation}
    \ket{\bm{k}\sigma} = \sum_{\bm{g}l}z_{\bm{k}\bm{g}l\sigma}\ket{\bm{k}+\bm{g},l,\sigma}.
\end{equation}
We give each value of moiré crystal momentum a unique representative $\bm{k}$ and refer to the set of these representatives as the moiré Brillouin zone ``mesh". We define the operator $[\;]$ to take an arbitrary momentum vector and replace it with its mesh equivalent such that $[\bm{k}+\bm{g}]=\bm{k}$ for all moiré reciprocal lattice vectors $\bm{g}$ where $\bm{k}$ is a mesh vector. Additionally, we define the reciprocal lattice (RL) part of an arbitrary momentum vector $\bm{g}_{\bm{q}} \equiv \bm{q}-[\bm{q}]$ so that $\bm{q}=[\bm{q}]+\bm{g}_{\bm{q}}$. From these Bloch states the interaction matrix elements can be calculated given the Fourier transform in two dimensions $V(\bm{q})$ of an interaction potential:
\begin{align}
    \begin{split}
    \bra{\bm{k}_1\sigma_1;\bm{k}_2\sigma_2}\hat{V}\ket{\bm{k}_3\sigma_3;\bm{k}_4\sigma_4}
    &= \frac{1}{A}\sum_{\bm{q}}V(\bm{q})\bra{\bm{k}_1\sigma_1}e^{i\bm{q}\cdot\bm{r}_1}\ket{\bm{k}_3\sigma_3}\bra{\bm{k}_2\sigma_2}e^{-i\bm{q}\cdot\bm{r}_2}\ket{\bm{k}_4\sigma_4} \\
    &= \frac{1}{A} \sum_{\bm{q}}V(\bm{q})\delta_{\bm{k}_1, [\bm{k}_3+\bm{q}]}\delta_{\bm{k}_2,[\bm{k}_4-\bm{q}]}\delta_{\sigma_1, \sigma_3}\delta_{\sigma_2, \sigma_4}\\
    &\times F(\bm{k}_1, \bm{k}_3,\sigma_3,\bm{g}_{\bm{k}_3+\bm{q}})F(\bm{k}_2, \bm{k}_4,\sigma_4,\bm{g}_{\bm{k}_4-\bm{q}})
    \end{split}
\end{align}
where
 \begin{align}
     F(\bm{k}_1, \bm{k}_2, \sigma, \bm{g}) = \sum_{\bm{g}'l}z^*_{\sigma\bm{k}_1(\bm{g}'+\bm{g})l}z_{\sigma\bm{k}_2\bm{g}'l}.
 \end{align}
$A$ is the system area. The Fourier transform in 2D of the Coulomb potential is $V(\bm{q})=\frac{2\pi e^2}{\epsilon |\bm{q}|}$. We neglect finite interlayer separation because it is much smaller than a typical moiré period. 
$\bm{q}$ is any momentum vector in the momentum space grid which is uniquely written as a sum of a mesh vector $\bm{k}$ and a moiré reciprocal lattice vector $\bm{g}$.
\par Since the reciprocal lattice is infinite even in a finite system, it must be truncated in practice. We truncate the reciprocal lattice vectors by retaining the $N$ shells of the lowest magnitude where an RL shell is a set of RL vectors of equal magnitude. Another natural possible choice is to retain all RL vectors $\bm{g}=n_1\bm{b}_1+n_2\bm{b}_2$ where $\bm{b}_1$, $\bm{b}_2$ are primitive RL vectors and $|n_i|\leq N$. As long as $N$ is sufficiently large, either scheme works. In practice, we use $N=30$, but this is far in excess of what is necessary to ensure convergence within numerical precision. Note that, in calculating the interaction matrix elements, one should include momentum transfers $\bm{q}$ whose RL component $\bm{g}_{\bm{q}}=\bm{g}'-\bm{g}$ for any $\bm{g}$, $\bm{g}'$ on the truncated RL, even when $\bm{g}'-\bm{g}$ is not on the truncated RL itself. However, the error incurred in not doing so is controlled by $N$, so it is not strictly necessary if $N$ is sufficiently large.
\par The $\bm{q}=0$ Fourier coefficient of the Coulomb interaction is divergent. To circumvent this issue, we send $V(\bm{q}=0)\rightarrow 0$. The contribution to the total energy arising from the $\bm{q}=0$ component of the particle-particle interaction is ${N_p\choose2}V(0) =\frac{N_p(N_p-1)V(0)}{2V}$ where $N_p$ is the number of particles and $V$ is the system volume in the appropriate spatial dimensions. Introducing a neutralizing, uniform charge background gives an additional $\bm{q}=0$ energy (accounting for both the positive energy contribution due to background-background interactions and the negative energy contribution coming from background-particle interactions) $+\frac{N_p^2}{2V}V(0)-\frac{N_p^2}{V}V(0)=-\frac{N_p^2}{2V}V(0)$. This cancels $\mathcal{O}(N_p^2)$ part of the particle-particle interaction, leaving a net energy contribution from the $\bm{q}=0$ component of the interaction $-\frac{N_p}{2V}V(0)$. The reason why the uniform background doesn't lead to perfect cancellation of the $\bm{q}=0$ energy is that the particles do not experience a self-interaction but the uniform background does. For a fixed $\frac{N_p}{V}$, the $\bm{q}=0$ energy is fixed and therefore becomes negligible in the thermodynamic limit, allowing the $\bm{q}=0$ component of the interaction to be dropped without consequence. In a finite system however the net $\bm{q}=0$ energy is finite but importantly depends only on $\frac{N_p}{A}$. All many-body energies reported in this work are energy differences between states at a fixed particle number and system area and are thus not affected by dropping the $\bm{q}=0$ part of the interaction. In fact, since the $\bm{q}=0$ energy only depends on $N_p$ and $A$ even without a neutralizing background, introducing a neutralizing background is really just a conceptual convenience in a finite size system since we end up neglecting the $-\frac{N_p}{2V}V(0)$ term anyway. For further discussion see, for instance, the first chapter of Ref. \cite{giuliani2005quantum}.
\par We define cluster geometries by the two vectors $\bm{L}_i$ under which the torus has periodic boundary conditions $\bm{r}=\bm{r}+\bm{L}_i$. All allowed plane waves on the torus are integer linear combinations of the basis plane waves $\bm T_i = \frac{2\pi \epsilon_{ij}\bm L_j \times \hat{z}}{A}$ which obey $\bm{T}_i\cdot \bm{L}_j=2\pi\delta_{ij}$. Here $A$ is the system area satisfying $\bm{L}_1\times \bm{L}_2=A\hat{z}$. Importantly, it is not necessary that $\bm{L}_i=N_i\bm{a}_i$ where $\bm{a}_i$ is a primitive moiré lattice vector and $N_i$ is an integer. For instance, the $\bm{L}_i$ of the 27-unit-cell cluster used in this work
are not of this form: $\bm{L}_1^{(27)}=3(2\bm{a}_1-\bm{a}_2)$, $\bm{L}_2^{(27)}=3(2\bm{a}_2-\bm{a}_1)$ with $\bm{a}_1=a_M(\frac{\sqrt{3}}{2},\frac{1}{2})$, $\bm{a}_2=a_M(0,1)$ in cartesian coordinates. Computationally, we find that it is best to represent all momentum vectors as integer vectors in the $\bm{T}_i$ basis. This avoids the issue of numerical instability in checking for equality between vectors that would come with storing them as float vectors in cartesian or primitive reciprocal lattice vector $\bm{b}_i$ bases.
\subsection{Computational methods}
\par We now summarize our computational methods for the many-body diagonalization. We write all calculations in the Julia programming language. We first construct a many-body basis of Fock states from the single-particle Bloch states. We assign each single particle Bloch state an integer label $i$ ranging from $0$ to the total number of Bloch states. We also assign the many-body Fock states are an integer label $I=\sum_{i \in occ}2^{i}$. Here ``$occ$" is the set of single-particle states that are occupied in a given Fock state. The $i^{th}$ entry of the bit string corresponding to the integer $I$ is $1$ if $i\in occ$ and $0$ otherwise. We compute the Hamiltonian matrix elements in the Fock state basis using creation and anhilation operator functions. These functions take in a Fock state integer $I$ and single particle state integer $i$. They return the Fock state integer $J$ such that $c^{\dag/()}\ket{I}=s\ket{J}$ where $s$ is a sign determined by Fermi statistics and that must be kept track of to determine the sign of the Hamiltonian matrix elements correctly. The creation and annihilation operator functions operate entirely through logical bit string manipulations on the integer representations $I$ of the Fock states. Particle number, $S_z$, and crystal momentum conservation means that the blocks of the Hamiltonian corresponding to a given many-body subspace with a fixed $N_p$ (=number of particles), $S_z$, and $\bm{k}_{tot}=[\sum_{i\in occ}\bm{k}_i]$ can be separately constructed and diagonalized. All many-body matrix elements in a given symmetry block are stored in a SparseMatrixCSC from the Sparse Arrays package in Julia. To diagonalize this sparse array, we use the \href{https://arpack.julialinearalgebra.org/stable/}{Arpack} package. We note anecdotally that, since performing the calculations shown in this work, we have found the \href{https://jutho.github.io/KrylovKit.jl/v0.5/}{KrylovKit} package to have higher performance for our purposes generally. We refer the reader to Ref. \cite{jafari2008introduction} for a pedagogical introduction to exact diagonalization techniques for many-fermion systems.

\subsection{Extended data}
\begin{figure*}
    \centering
\includegraphics[width=\textwidth]{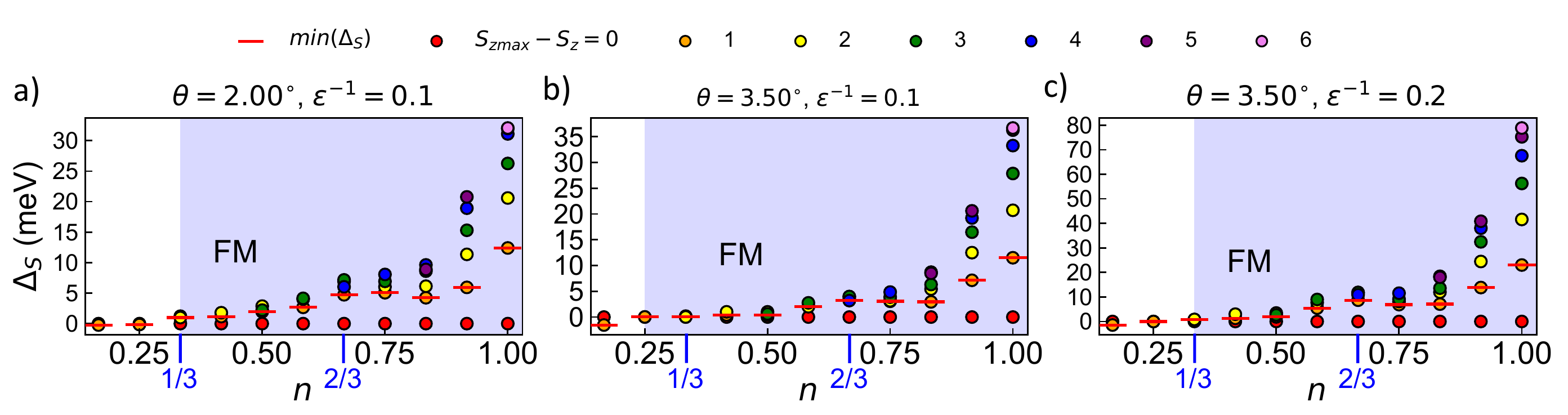}
    \caption{$\Delta_{S}\equiv E_{\text{min}}(S_z)-E_{\text{min}}(S_{z\text{max}})$ across all possible $S_z$ values as a function of the filling factor $n\equiv N_h/N_{uc}$ on the 12-unit-cell cluster at several $\epsilon^{-1}$, $\theta$. min($\Delta_S$) is the minimum value of $\Delta_S$ for $S_z < S_{z\text{max}}$.} 
\label{fig:extendedmagnetism}
\end{figure*}
\begin{figure*}
    \centering
\includegraphics[width=0.6\textwidth]{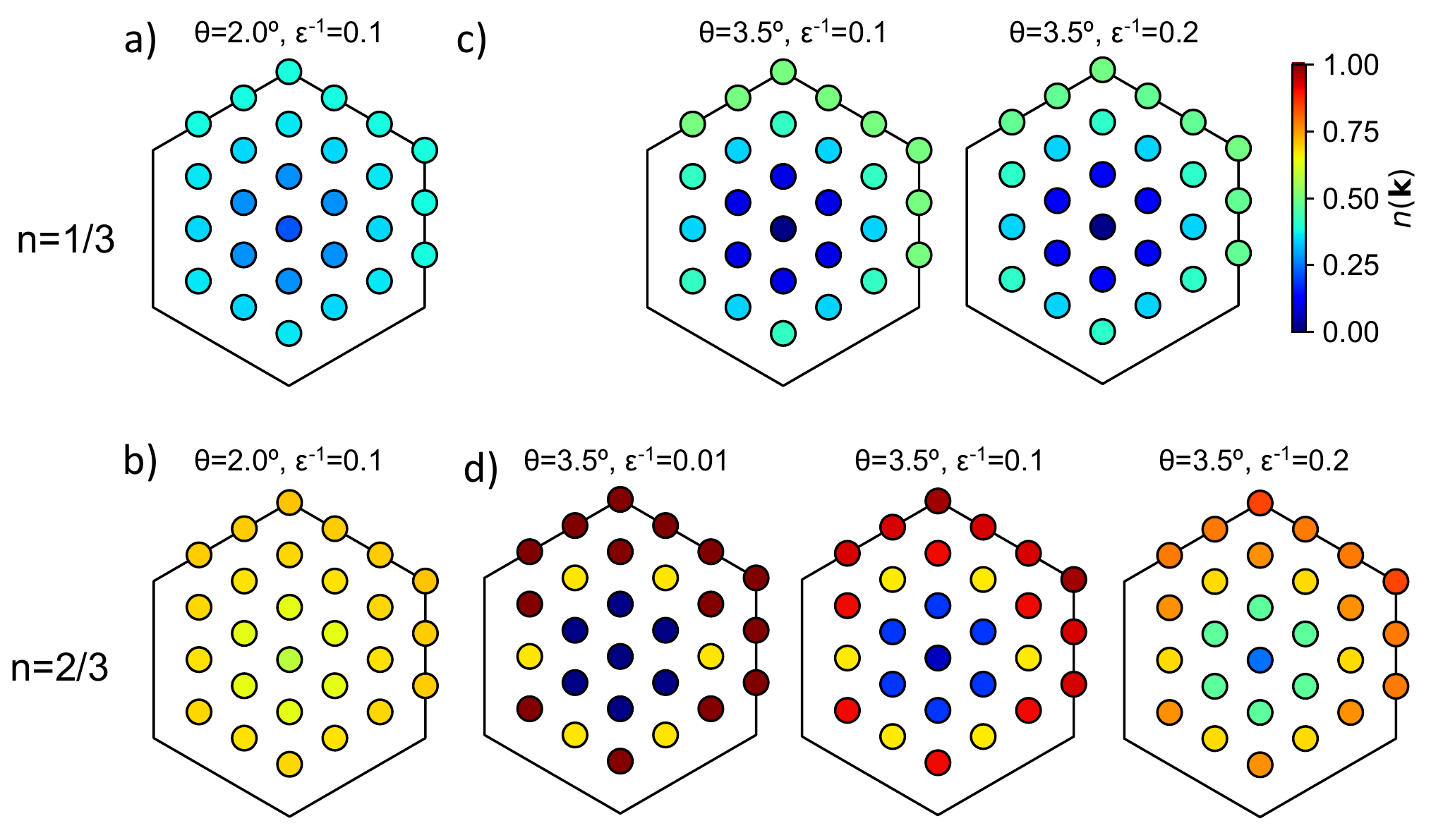}
    \caption{Bloch state occupation numbers $n(\bm{k})=\bra{\Psi}c^{\dag}_{\bm{k}}c_{\bm{k}}\ket{\Psi}$ where $\Psi$ is the many-body ground state at several twist angles and interaction strengths and filling factors $n=\frac{1}{3}$ (top row) and $\frac{2}{3}$ (bottom row).} 
\label{fig:blochocc}
\end{figure*}
\begin{figure*}
    \centering
\includegraphics[width=0.6\textwidth]{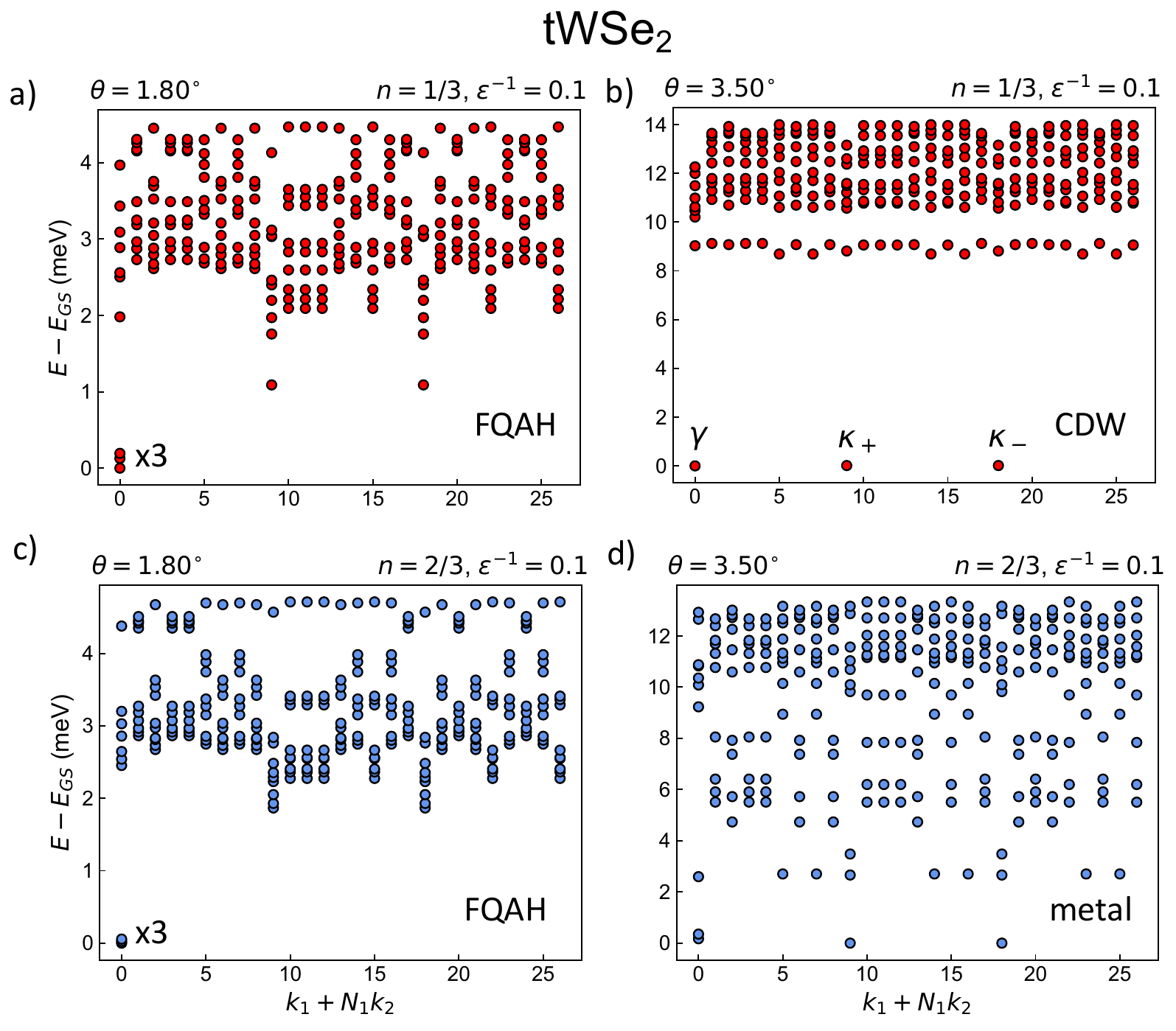}
    \caption{Low-lying many-body spectra of $t$WSe$_2$ at filling factors $n=\frac{1}{3}$, $\frac{2}{3}$ at a fixed interaction strength $\epsilon^{-1}=0.1$ and two twist angles $\theta=1.80^{\circ}$, $3.50^{\circ}$. Full valley polarization is assumed.} 
\label{fig:wse2}
\end{figure*}
In Fig. \ref{fig:extendedmagnetism}, we show additional data on magnetism with respect to doping to supplement the data presented in Fig. 6 of the main text but from a different cluster defined by $\bm{L}^{(12)}_1=2(2\bm{a}_1-\bm{a}_2)$, $\bm{L}^{(12)}_2=2(2\bm{a}_2-\bm{a}_1)$. This cluster contains $12$ unit cells and includes the $\gamma$, $m$, and $\kappa_{\pm}$ points of the moiré Brillouin zone. While this is smaller than the $15$-unit-cell cluster used in Fig. 6 of the main text, it has the advantage that it better respects the moiré superlattice point group symmetries.
\par In Fig. \ref{fig:blochocc}, we present additional evidence for various phases occurring at filling factors $n=\frac{1}{3}$ and $\frac{2}{3}$ at different twist angles. Throughout Fig. \ref{fig:blochocc} we assume full spin/valley polarization. Fig. \ref{fig:blochocc}(a,b) shows the Bloch state occupation numbers $n(\bm{k})$ of the many-body ground states at $\theta=2^{\circ}$. $n(\bm{k})=\bra{\Psi}c^{\dag}_{\bm{k}}c_{\bm{k}}\ket{\Psi}$ where $\Psi$ is the many-body ground state which, in all cases shown in Fig. \ref{fig:blochocc} has many-body crystal momentum $\gamma$. In both cases the Bloch state occupation is nearly uniform, consistent with an insulating FQAH ground state.
\par In Fig. \ref{fig:blochocc}(c), we show $n(\bm{k})$ at $\theta=3.5^{\circ}$, $n=\frac{1}{3}$ for moderate and strong interactions $(\epsilon^{-1}=0.1, 0.2)$. While $n(\bm{k})$ is not completely uniform, it lacks a clear Fermi surface, consistent with our claim in the main text that this state is a charge density wave (CDW). The similarity between $n(\bm{k})$ in the presence of moderate and strong interactions also supports our claim that the system at this filling factor is already effectively in the strongly-interacting limit at $\epsilon^{-1}=0.1$. In Fig. \ref{fig:blochocc}(d), we show $n(\bm{k})$ at $\theta=3.5^{\circ}$, $n=\frac{2}{3}$ for weak, moderate, and strong interactions $(\epsilon^{-1}=0.01,0.1, 0.2)$. When interactions are weak, $n(\bm{k})$ exhibits a clear Fermi surface across which $n(\bm{k})$ drops sharply. The partial occupation of some Bloch states is a finite-size effect resulting from the degeneracy of several configurations in the absence of interactions. At intermediate interaction strength, the Fermi surface is less sharp but still present, consistent with our claim in the main text that this state is a ferromagnetic metal. At strong interaction strength, $n(\bm{k})$ is more uniform and lacks a Fermi surface, consistent with an insulating FQAH state.
\par In Fig. \ref{fig:wse2}, we present many-body spectra for $t$WSe$_2$, demonstrating that the same phases we study in $t$MoTe$_2$ can also occur in this system. For an explanation of how we infer the various phases corresponding to the many-body spectra shown in Fig. \ref{fig:wse2}, see the discussion surrounding Figs. (7-9) of the main text. We use the continuum model parameters listed in Table 1 of the main text.

\section{Moiré band topology from continuum model symmetry analysis}
We seek to determine moiré band's Chern number in the nearly-free limit
\begin{align}
    |V|, |w| \ll \frac{\hbar^2}{2ma_M^2}
\end{align}
of the continuum model realized at large twist angle. To do so, we build off of analyses presented in the supplements of Refs. \cite{paul2023giant,devakul2021magic}. For simplicity, we focus our analysis on the $K$ valley.
\par We start from the following equation relating the $C_{3z}$ eigenvalues of the high-symmetry point Bloch states of a given moiré band to the moiré band's Chern number $C$ (see Eq. 32 of Ref. \cite{fang2012bulk}):
\begin{align}\label{eq:chernmod3}
C \mod 3 = \frac{3}{2\pi}\arg(- \lambda_{\kappa_+} \lambda_{\kappa_-} \lambda_{\gamma}).
\end{align}
Here $\mathcal{R}_{\theta}=e^{-i\frac{\theta}{\hbar}\hat{J}_z} = e^{-i\frac{\theta}{2}\sigma_z}e^{-i\frac{\theta}{\hbar}\hat{L}_z}$ is a counter-clockwise rotation operator about the $z$-axis for spin$-\frac{1}{2}$ fermions and $\lambda_{\bm k} = \bra{\psi_{\bm k}}\mathcal{R}_{2\pi/3}\ket{\psi_{\bm k}}$ are $C_{3z}$ symmetry eigenvalues at three moiré Brillouin zone high symmetry points ${\bm k}=\kappa_+, \kappa_-, \gamma$. The origin with respect to which the rotation is defined is the center of an MM stacking region where a metal atom in the top layer is directly above a metal atom in the bottom layer.
\par The eigenstates of the ``kinetic" part of the continuum model Hamiltonian (i.e. the part present when $w=V=0$) are monolayer Bloch states near the $K$ valley of each layer. We can label these states with monolayer crystal momentum measured relative to the moiré Brillouin zone $\gamma$ point and a layer index $\ket{\bm{k},l}$ (we have dropped by spin/valley label because it is assumed throughout this section that we are dealing with valley $K$). These states have unperturbed energies $\varepsilon_{\bm{k},1}=\frac{\hbar^2|\bm{k}-\kappa_+|^2}{2m}$, $\varepsilon_{\bm{k},2}=\frac{\hbar^2|\bm{k}-\kappa_-|^2}{2m}$. Our Brillouin zone folding scheme is such that $K_{l}\rightarrow \kappa_l$ where $K_l$ is a monolayer $K$ point and $\kappa_{1,2} = \kappa_{+,-}$. Let's first understand how these states transform under $\mathcal{R}_{\frac{2\pi}{3}}$. For this purpose it is helpful to define monolayer Bloch states with momentum measured relative to the monolayer $\Gamma$ point, $\ket{\bm{k},l}=\ket{\bm{k}+K_l-\kappa_l,l}_{ML}=e^{i\bm{k}\cdot\bm{r}}\ket{u_{\bm{k}},l}_{ML}$. The continuum model makes the approximation $\ket{u_{K_l+\bm{\delta}},l}=\ket{u_{K_l},l}$ which is accurate when $\bm{\delta}$ is sufficiently small. Since $K_l$ is a high symmetry point with respect to $C_{3z}$ (meaning that $\mathcal{R}_{\frac{2\pi}{3}}K_l=K_l+\bm{G}_l$ where $\bm{G}_l$ is a monolayer reciprocal lattice vector), and each monolayer has $C_{3z}$ symmetry, $\ket{K_l,l}_{ML}$ is an eigenstate of $\mathcal{R}_{\frac{2\pi}{3}}$. Let's define its eigenvalue $\mathcal{R}_{\frac{2\pi}{3}}\ket{K_l,l}_{ML}=\lambda_{K_l,l}\ket{K_l,l}_{ML}$. Since both layers are identical we have $\lambda_{K_1,1}=\lambda_{K_2,2} \equiv \lambda_K$. From the continuum model approximation $\ket{u_{K_l+\bm{\delta}},l}=\ket{u_{K_l},l}$ it follows that
\begin{align}
\begin{split}
      \mathcal{R}_{\frac{2\pi}{3}}\ket{K_l+\bm{\delta},l}_{ML} &=   \mathcal{R}_{\frac{2\pi}{3}}e^{i\bm{\delta}\cdot\bm{r}}\ket{K_l,l}_{ML} \\
      &= e^{i\mathcal{R}_{\frac{2\pi}{3}}\bm{\delta}\cdot\bm{r}}\mathcal{R}_{\frac{2\pi}{3}}\ket{K_l,l}_{ML} \\
      &=e^{i\mathcal{R}_{\frac{2\pi}{3}}\bm{\delta}\cdot\bm{r}}\lambda_{K_l,l}\ket{K_l,l}_{ML} \\
      &= \lambda_{K_l,l}\ket{K_l+\mathcal{R}_{\frac{2\pi}{3}}\bm{\delta},l}_{ML} 
\end{split}
\end{align}
\par 
This tells us that layer Bloch states with their momentum measured relative to the moiré Brillouin zone $\gamma$ point transform as
\begin{align}
    \begin{split}
        \mathcal{R}_{\frac{2\pi}{3}}\ket{\bm{k},l} &= \mathcal{R}_{\frac{2\pi}{3}}\ket{\bm{k}+K_l-\kappa_l,l}_{ML} \\
        &= \lambda_{K}\ket{K_l+\mathcal{R}_{\frac{2\pi}{3}}(\bm{k}-\kappa_l),l}_{ML} \\
        &= \lambda_{K}\ket{K_l+\mathcal{R}_{\frac{2\pi}{3}}(\bm{k}-\kappa_l),l}_{ML} \\
        &= \lambda_{K}\ket{\kappa_l+\mathcal{R}_{\frac{2\pi}{3}}(\bm{k}-\kappa_l),l}.
    \end{split}
\end{align}
In other words, we should rotate the $\bm{k}$ in $\ket{\bm{k},l}$ about $\kappa_l$, \emph{not} about $\gamma$.
\par When $V=w=0$, there are six degenerate layer Bloch states that fold to $\gamma$ of the moiré Brillouin zone: $\ket{1}=\ket{\gamma,1}$,
$\ket{2}=\ket{\bm{g}_6,2}$,
$\ket{3}=\ket{\bm{g}_1,1}$,
$\ket{4}=\ket{\bm{g}_1,2}$, $\ket{5}=\ket{\bm{g}_2,1}$,   and $\ket{6}=\ket{\gamma,2}$.
The continuum model Hamiltonian acts on these states as 
\begin{align}
\begin{split}\label{eq:degenham}
    \mathcal{H}\ket{j} = \frac{\hbar^2|\kappa_+|^2}{2m}\ket{j} -V(e^{-i\phi}\ket{j+2}+e^{+i\phi}\ket{j-2})+w(\ket{j+1}+\ket{j-1}) +\ldots
\end{split}
\end{align}
where $\dots$ denotes the part of $\mathcal{H}\ket{j}$ outside of the degenerate subspace.
Here we identify $\ket{j}=\ket{j\pm 6}$. It follows from Eq. \ref{eq:degenham} that $\mathcal{P}[P_{+1},\mathcal{H}]\mathcal{P}=0$ where $\mathcal{P}$ is the projector onto the degenerate subspace and $P_{+1}\ket{j}=\ket{j+1}$. Therefore, $P_{+1}$ and $\mathcal{H}$ are simultaneously diagonalizable within the degenerate subspace and the eigenstates take the form
\begin{align}
    \begin{split}
        \ket{\psi_n} = \frac{1}{\sqrt{6}}\sum_{j=1}^{6}e^{i\frac{jn\pi}{3}}\ket{j}
    \end{split}
\end{align}
so that $P_{+1}\ket{\psi_n}=e^{-i\frac{n\pi}{3}}\ket{\psi_n}$
with energy eigenvalues
\begin{align}
    \begin{split}
        \varepsilon_n = \frac{\hbar^2|\kappa_+|^2}{2m}-2V\cos(\frac{n2\pi}{3}+\phi)+2w\cos(\frac{n\pi}{3}).
    \end{split}
\end{align}
Applying the rotation rule derived above to the states $\ket{j}$ gives
\begin{align}
\begin{split}
        \mathcal{R}_{\frac{2\pi}{3}}\ket{j} &= \lambda_K\ket{j+2} \\
        &= \lambda_K (P_{+1})^2\ket{j}
\end{split}
\end{align}
which implies that
\begin{align}
\begin{split}
    \lambda_{n} &= \bra{\psi_{n}}\mathcal{R}_{\frac{2\pi}{3}}\ket{\psi_n}\\
    &= \lambda_{K} e^{-in\frac{2\pi}{3}}.
\end{split}
\end{align}
\par From the continuum model Hamiltonian it is apparent that $\kappa_{+}/\kappa_{-}$ is the minimum of the parabolic dispersion in layer $1$/$2$. The layer Bloch states at these points are not degenerate with any other states to which they are related by a moiré reciprocal lattice vector. Therefore, at leading order in $V,\,w$, the lowest moiré band Bloch states at the Brillouin zone corners are simply $\ket{\kappa_+}=\ket{\kappa_+,1}=\ket{K_1,1}_{ML}$, $\ket{\kappa_-}=\ket{\kappa_-,2}=\ket{K_2,2}_{ML}$ with symmetry eigenvalues $\lambda_{K}$.
\par To produce Fig. 4b of the main text, we determine $n_{min}$ such that $\varepsilon_{n_{min}}$ is the lowest among the $\varepsilon_n$ as a function of $\frac{w}{V}$ and $\phi$. We then determine the Chern number of the lowest moiré band in valley $K$ mod $3$ by plugging $\lambda_{\gamma}=\lambda_{n_{min}}$ into Eq. \ref{eq:chernmod3}: $C \mod 3 = \frac{3}{2\pi}\arg(- \lambda_{\kappa_+} \lambda_{\kappa_-} \lambda_{\gamma})= \frac{3}{2\pi}\arg(-(\lambda_{K})^3e^{-i\frac{n_{min}2\pi}{3}}) =  (-n_{min}) \mod 3$ where we use $(\mathcal{R}_{\frac{2\pi}{3}})^3=(e^{-i\frac{\pi}{3}\sigma_z}e^{-i\frac{2\pi}{3\hbar}\hat{L}_z})^3=e^{-i\pi\sigma_z}e^{-i2\pi\hat{L}_z}=-1 \implies \lambda_{K}^3=-1$.
\section{Moiré band quantum geometry}
\begin{figure*}[h]
    \centering
\includegraphics[width=0.6\textwidth]{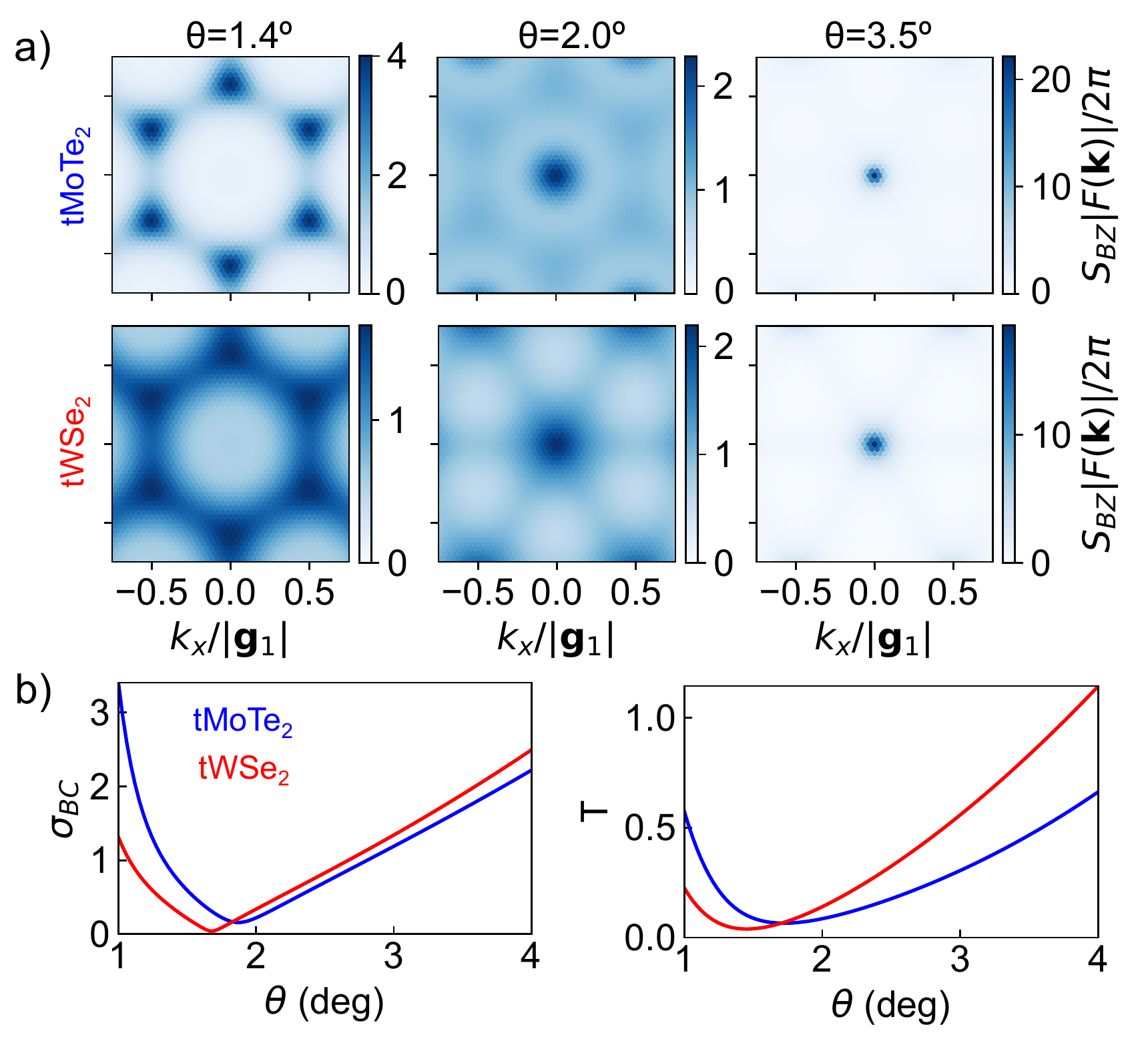}
    \caption{(a) Absolute value of the Berry curvature of the lowest moiré bands of $t$MoTe$_2$ and $t$WSe$_2$ at several twist angles. (b) Berry curvature standard deviation $\sigma_{BC}$ and trace condition violation $T$ of the two systems' lowest moiré bands as a function of twist angle. $S_{BZ}=|\bm{b}_1\times\bm{b}_2|$ is the moiré Brillouin zone momentum space area.} 
\label{fig:quantumgeometry}
\end{figure*}

A growing body of work has demonstrated connections between the quantum geometry of a given Chern band and its propensity to host FQAH states. We present calculations of Berry curvature and trace condition violation of our system in Fig. \ref{fig:quantumgeometry}. Both $t$MoTe$_2$ and $t$WSe$_2$ exhibit deep minima in Berry curvature standard deviation
\begin{align}
   \sigma_{BC} = \left(\int_{BZ} d^2\bm{k}\left[\frac{F(\bm{k})}{2\pi}-\frac{C}{S_{BZ}}\right]^2\right)^{1/2}
\end{align}
and trace condition violation
\begin{align}
   T = \left(\int_{BZ} d^2\bm{k}\frac{\tr g(\bm{k})}{2\pi}\right) - |C|
\end{align}
where $F(\bm{k})$ and $g(\bm{k})$ are the Berry curvature and Fubini-Study metric respectively
in the approximate twist angle range $1.4^{\circ}\leq \theta\leq 1.9^{\circ}$. Our finding that the FQAH gap is maximized near $\theta\approx 2^{\circ}$ agrees reasonably well with the statement that low $T$ and $\sigma_{BC}$ are favorable conditions for FQAH states. (Note that our definition of $T$ differs by a factor of $\frac{1}{2\pi}$ from some definitions given elsewhere.)




%